\documentclass[journal]{IEEEtran}
\usepackage{blindtext}
\usepackage[latin1]{inputenc}
\usepackage[noadjust]{cite}
\usepackage{graphicx}
\usepackage{stfloats}
\usepackage[font=footnotesize]{caption}
\usepackage{psfrag}
\usepackage{subfig}
\usepackage{amsmath,amssymb,amsthm,mathrsfs,amsfonts,dsfont}
\usepackage{color}
\usepackage[algoruled]{algorithm2e}
\usepackage{fixltx2e}
\usepackage{multirow}
\usepackage{multicol}
\usepackage[normalem]{ulem}
\usepackage{acronym}
\usepackage[table,xcdraw]{xcolor}
\usepackage{accents}
\usepackage{mwe}
\usepackage{hhline}
\usepackage{trfsigns}
\usepackage{siunitx}
\usepackage{slashbox}
\usepackage{wasysym}

\newacro{5G}{fifth generation}
\newacro{6G}{sixth generation}
\newacro{A/D}{analog-to-digital}
\newacro{ADC}{analog-to-digital converter}
\newacro{AWGN}{additive white Gaussian noise}
\newacro{BB}{baseband}
\newacro{BER}{bit error ratio}
\newacro{BPSK}{binary phase-shift keying}
\newacro{BP}{band-pass}
\newacro{CCDF}{complementary cumulative distribution function}
\newacro{CDM}{code-division multiplexing}
\newacro{CFO}{carrier frequency offset}
\newacro{CFR}{channel frequency response}
\newacro{CIR}{channel impulse response}
\newacro{CP}{cyclic prefix}
\newacro{CPO}{carrier phase offset}
\newacro{CR}[C\&R]{wireless communication and radar sensing}
\newacro{CS}{chirp sequence}
\newacro{CW}{continuous wave}
\newacro{D/A}{digital-to-analog}
\newacro{DAC}{digital-to-analog converter}
\newacro{DDS}{direct digital synthesis}
\newacro{DFCS}{dual-functional communication and radar sensing}
\newacro{DFRC}{dual-functional radar-communication}
\newacro{DFnT}{discrete Fresnel transform}
\newacro{DFT}{discrete Fourier transform}
\newacro{DMRS}{demodulation reference signal}
\newacro{DoA}{direction of arrival}
\newacro{EVM}{error vector magnitude}
\newacro{FBMC}{filter bank multicarrier}
\newacro{FDE}{frequency-domain equalization}
\newacro{FDM}{frequency-division multiplexing}
\newacro{FDZP}{frequency-domain zero padding}
\newacro{FMCW}{frequency-modulated continuous wave}
\newacro{FrDM}{Fresnel-division multiplexing}
\newacro{FSP}{frequency-shift precoding}
\newacro{GFDM}{generalized frequency-division multiplexing}
\newacro{HAD}{highly automated driving}
\newacro{HP}{high-pass}
\newacro{IBFD}{in-band full duplex}
\newacro{IC}{interference cancellation}
\newacro{ICI}{intercarrier interference}
\newacro{IChI}{interchirp interference}
\newacro{IDFT}{inverse discrete Fourier transform}
\newacro{IDFnT}{inverse discrete Fresnel transform}
\newacro{IF}{intermediate frequency}
\newacro{ISAC}{integrated sensing and communication}
\newacro{ISI}{intersymbol interference}
\newacro{ISLR}{integrated-sidelobe level ratio}
\newacro{JCAS}{joint communication and sensing}
\newacro{JRC}{joint radar-communications}
\newacro{LNA}{low-noise amplifier}
\newacro{LO}{local oscillator}
\newacro{LoS}{line-of-sight}
\newacro{LP}{low-pass}
\newacro{LFSR}{linear-feedback shift register}
\newacro{mmWave}{milimeter wave}
\newacro{MIMO}{multiple-input multiple-output}
\newacro{MLS}{maximum-length sequence}
\newacro{MMSE}{minimum mean square error}
\newacro{MU}{multi-user}
\newacro{NBI}{narrowband interference}
\newacro{NLoS}{non-line-of-sight}
\newacro{NR}{New Radio}
\newacro{NRMSE}{normalized root mean square error}
\newacro{OCDM}{orthogonal chirp-division multiplexing}
\newacro{OFDM}{orthogonal frequency-division multiplexing}
\newacro{OOB}{out-of-band}
\newacro{OTFS}{orthogonal time-frequency space}
\newacro{P/S}{parallel-to-serial}
\newacro{PA}{power amplifier}
\newacro{PACF}{periodic autocorrelation function}
\newacro{PAPR}{peak-to-average power ratio}
\newacro{PCCF}{periodic cross-correlation function}
\newacro{PLC}{powerline communication}
\newacro{PLL}{phase-locked loop}
\newacro{PMCW}{phase-modulated continuous wave}
\newacro{PMEPR}{peak-to-mean envelope power ratio}
\newacro{PPLR}{peak power loss ratio}
\newacro{PRBS}{pseudorandom binary sequence}
\newacro{PSLR}{peak sidelobe level ratio}
\newacro{QPSK}{quadrature phase-shift keying}
\newacro{RaaS}{radar as a service}
\newacro{RadCom}{radar-communication}
\newacro{RCS}{radar cross section}
\newacro{RF}{radio-frequency}
\newacro{RTS}{radar target simulator}
\newacro{SDM}{spatial division multiplexing}
\newacro{SDMA}{spatial division multiple access}
\newacro{SH}[S\&H]{sample and hold}
\newacro{SI}{self-interference}
\newacro{SIC}{self-interference cancellation}
\newacro{SISO}{single-input single-output}
\newacro{SNR}{signal-to-noise ratio}
\newacro{S/P}{serial-to-parallel}
\newacro{SC}[S\&C]{Schmidl \& Cox}
\newacro{SoC}{system-on-a-chip}
\newacro{TDD}{time-division duplexing}
\newacro{TDE}{time-domain equalization}
\newacro{TDM}{time-division multiplexing}
\newacro{TDR}{time-domain reflectometry}
\newacro{ToF}{time of flight}
\newacro{UAV}{unmanned aerial vehicle}
\newacro{UE}{user equipment}
\newacro{UWAC}{underwater acoustic communication}
\newacro{V2I}{vehicle-to-infrastructure}
\newacro{V2V}{vehicle-to-vehicle}
\newacro{ZCZ}{zero correlation zone}
\newacro{ZF}{zero forcing}
\newacro{ZP}{zero padding}

\makeatletter
\renewcommand*\env@cases[1][1.2]{%
	\let\@ifnextchar\new@ifnextchar
	\left\lbrace
	\def\arraystretch{#1}%
	\array{@{}l@{\quad}l@{}}%
}
\makeatother
\usepackage{fancyhdr}
\fancypagestyle{empty}{fancy}

\begin{document}
	
%\title{Fresnel-Domain Processing for Improved Doppler-Shift Robustness in OCDM Radar Systems}
%\title{On the Influence of Time and Frequency Shifts in Fresnel-Domain Channel Estimation in OCDM-based Systems}
%\title{On Fresnel-Domain Channel Estimation in OCDM-based Radar and Communication Systems}
\title{Discrete-Fresnel Domain Channel Estimation in OCDM-based Radar Systems}

\author{Lucas Giroto de Oliveira,~\IEEEmembership{Graduate Student Member,~IEEE},
	Benjamin Nuss,~\IEEEmembership{Graduate Student Member,~IEEE},
	Mohamad Basim Alabd,~\IEEEmembership{Graduate Student Member,~IEEE},
	Axel Diewald,~\IEEEmembership{Graduate Student Member,~IEEE},
	Yueheng Li,~\IEEEmembership{Graduate Student Member,~IEEE}, Linda Gehre, Xueyun Long,\\ Theresa Antes, Johannes Galinsky, and Thomas Zwick,~\IEEEmembership{Fellow,~IEEE}
	\thanks{Manuscript received MM DD, 2022. The authors acknowledge the financial support by the Federal Ministry of Education and Research of Germany in the project ``Open6GHub'' (grant number: 16KISK010). The work of Lucas Giroto de Oliveira was also financed by the German Academic Exchange Service (DAAD) - Funding program 57440921/Pers. Ref. No. 91555731. \textit{(Corresponding author: Lucas Giroto de Oliveira.)}}
	\thanks{The authors are with the Institute of Radio Frequency Engineering and Electronics (IHE), Karlsruhe Institute of Technology (KIT), 76131 Karlsruhe, Germany (e-mail: {lucas.oliveira@kit.edu}).}
	%\thanks{Color versions of one or more figures in this article are available at https://doi.org/10.1109/TMTT.2022.XXXXXXX.}
	%\thanks{Digital Object Identifier 10.1109/TMTT.2022.XXXXXXX}
}

%\markboth{IEEE Transactions on Microwave Theory and Techniques}
%{GIROTO DE OLIVEIRA \MakeLowercase{\textit{et al.}}: \uppercase{On Discrete-Fresnel Domain Channel Estimation in OCDM-based Radar and Communication Systems}}

\maketitle

\begin{abstract}
	In recent years, orthogonal chirp-division multiplexing (OCDM) has been increasingly considered as an alternative multicarrier scheme, e.g., to orthogonal frequency-division multiplexing, in digital communication applications. Among reasons for thar are its demonstrated superior performance resulting from its robustness to impairments such as frequency selectivity of channels and intersymbol interference. Furthermore, the so-called unbiased channel estimation in the discrete-Fresnel domain has also been investigated for both communication and sensing systems, however without considering the effects of frequency shifts. This article investigates the suitability of the aforementioned discrete-Fresnel domain channel estimation in OCDM-based radar systems as an alternative to the correlation-based processing previously adopted, e.g., in the radar-communication (RadCom) literature, which yields high sidelobe level depending on the symbols modulated onto the orthogonal subchirps. In this context, a mathematical formulation for the aforementioned channel estimation approach is introduced. Additionally, extensions to multi-user/multiple-input multiple-output and RadCom operations are proposed. Finally, the performance of the proposed schemes is analyzed, and the presented discussion is supported by simulation and measurement results. In summary, all proposed OCDM-based schemes yield comparable radar sensing performance to their orthogonal frequency-division multiplexing counterpart, while achieving improved peak-to-average power ratio and, in the RadCom case, communication performance.%Finally, the performance of the proposed schemes is analyzed, and the presented discussion is supported by simulation and measurement results.
\end{abstract}


\IEEEpeerreviewmaketitle

%***********************************************************%
%-----------------------------------------------------------%
%***********************************************************%

\section{Introduction}\label{sec:introduction}

\IEEEPARstart{A}{ll}-digital radar systems have been increasingly gaining attention in recent years. While they impose the challenge of handling high data rates resulting from the use of  \acp{DAC} and \acp{ADC} with high resolutions and sampling rates \cite{schweizer2021}, their use enables a wide range of possibilities. Among these are enabling efficient \ac{MU} or \ac{MIMO} operation, e.g., for distributed radar sensing or \ac{DoA} estimation \cite{vasanelli2020}, while also yielding high unambiguous velocity and fine range resolution. Further aspects of all-digital radar systems also include higher signal processing flexibility and improved performance for joint \ac{RadCom} operation \cite{giroto2021_tmtt}.

The possibilities enabled by all-digital radar systems rely on the use of efficient modulation schemes, among which are the widely-known \ac{OFDM}, \ac{PMCW} \cite{hakobyan2019,roos2019,waldschmidt2021,giroto2021_mwcl,giroto2022_PMCW_MDPI}, and \ac{OCDM} \cite{ouyang2016}. The latter has been recently investigated in an \ac{ISAC} context \cite{wang2021,liu2022} to enable joint \ac{RadCom} operation  \cite{giroto2020_OCDM_RadCom,giroto2021_OCDM_RadCom} or applications such as joint sonar-communication \cite{wang2022}. On the one hand, \ac{OCDM} is known to outperform its aforementioned counterparts for communication purposes \cite{giroto2021_tmtt,omar2021}, e.g., due to its multicarrier and spread spectrum characteristics \cite{ouyang2015,ouyang2017} that yield robustness to multipath propagation, Doppler shifts, \ac{ISI} \cite{bomfin2019}, and \ac{NBI} \cite{omar2020_NBI}. On the other hand, the radar sensing performance of conventional \ac{OCDM}-based systems is limited. Examples include the relatively high range sidelobe level that is influenced by modulated symbols onto the orthogonal subchirps in the \ac{OCDM}-based \ac{RadCom} systems with correlation-based radar signal processing \cite{giroto2020_OCDM_RadCom,giroto2021_OCDM_RadCom,giroto2021_tmtt,bhattacharjee2021}.
%\cor{\IEEEPARstart{A}{b} \blindtext}

In the face of limitations of typical radar processing schemes for \ac{OCDM}-based systems, a potential alternative is the use of discrete-Fresnel domain channel estimation, which has been introduced in \cite{ouyang2017_unbiasedCE} for coherent optical \ac{OFDM} systems and further investigated in \cite{ouyang2018_unbiasedCE} and \cite{zhang2022}. The aforementioned channel estimation strategy has also been used in the context of sensing, namely for enabling reflectometric sensing of power lines \cite{giroto2021} with a baseband \ac{OCDM} system \cite{dib2020}. The simplest discrete-Fresnel domain channel estimation strategy consists of transmitting a pilot \ac{OCDM} symbol where only the first subchirp is active. Based on the convolution theorem of the \ac{DFnT} \cite{ouyang2015}, the receive \ac{OCDM} symbol will ultimately contain a \ac{CIR} estimate. The aforementioned studies have, however, considered that the estimated channel only introduces delays in the transmit signal. Frequency shifts were either not considered \cite{ouyang2017_unbiasedCE,ouyang2018_unbiasedCE} or assumed to be compensated \cite{zhang2022}. While the assumption that frequency shifts are compensated may be reasonable for communication purposes, where, e.g., Schmidl \& Cox's algorithm \cite{schmidl1997,ouyang2017_diss} can be used, it does not necessarily hold for radar systems. Although transmit and receive channels usually share the same oscillator and are fully synchronized, Doppler shifts introduced by the motion of targets are not compensated during radar channel estimation. In fact, the goal of radar sensing is jointly estimating the range and Doppler shift, and consequently relative radial velocity, of targets. The lack of compensation for Doppler shifts, which are essentially frequency shifts, results in the need for a more thorough investigation of the discrete-Fresnel domain channel estimation for \ac{OCDM}-based radar systems, which is the subject of this article.

%In this context, the main contributions of this article are as follows:
The main contributions of this article are as follows:
\begin{itemize}
	
	\item A thorough mathematical formulation of the effects of time and frequency shifts in the discrete-Fresnel domain, both applied individually or jointly. It is shown that both aforementioned effects cause shifts and phase rotations of subchirps in the discrete-Fresnel domain, leading to \ac{IChI} and resembling the \ac{ICI} in \ac{OFDM}-based systems \cite{hwang2009}. These findings extend the state-of-the-art knowledge on propagation effects on \ac{OCDM} signals, which previously only consider the effects of time shifts described by the convolution theorem of the \ac{DFnT}, and can be used in the contexts of radar, communication, and \ac{RadCom} systems.
	
	\item A detailed description of discrete-Fresnel domain radar channel estimation, with closed-form expressions to analyze results for multiple point targets with distinct ranges and relative radial velocities and radar performance parameters including processing gain, range resolution and ambiguity, and velocity resolution and ambiguity.
	
	\item An extension of the investigated discrete-Fresnel domain radar channel estimation to \ac{MU}/\ac{MIMO} and \ac{RadCom} operation based on a strategic subchirp allocation known as \ac{FrDM}, with remarks on eventually changed radar performance parameters.
	
	\item A performance analysis based on simulation and measurement results. The presented analysis covers an investigation of biases from ideally estimated \acp{CIR} under different Doppler shifts, which is supported by an analysis of \ac{SNR} degradation and sidelobe level changes. Additionally, comparisons with the well-known \ac{OFDM} \ac{RadCom} are performed. The achieved results show that the proposed \ac{OCDM}-based schemes present comparable radar sensing performance to \ac{OFDM}, while presenting lower \ac{PAPR} due to a similar effect to the known tone reservation in \ac{OFDM} systems. Furthermore, the proposed extension of the proposed \ac{OCDM}-based radar system to \ac{RadCom} operation, which is named sector-modulated \ac{OCDM}-based \ac{RadCom},  yields improved communication robustness w.r.t. \ac{OFDM}, as it retains the spread-spectrum characteristics of conventional \ac{OCDM} schemes.%radar in terms of radar performance and \ac{PAPR} are performed, and the communication performance of the proposed \ac{OCDM}-based \ac{RadCom} system is assessed.
	
\end{itemize}

%\corb{\ac{OCDM}} - \cor{1st relevant paper and OCDM principles: \cite{ouyang2016,ouyang2015,ouyang2017,ouyang2017_diss}} - \cor{Channel estimation: \cite{ouyang2017_unbiasedCE,ouyang2018_unbiasedCE,giroto2021} and \cite{zhang2022}} - \cor{Communication:} - \cor{ISAC: \cite{giroto2020_OCDM_RadCom,giroto2021_OCDM_RadCom,giroto2021_tmtt}}

The remainder of this article is organized as follows. Section~\ref{sec:time-freqConv} presents a thorough mathematical description of the individual and joint effects of time and frequency shifts in the discrete-Fresnel domain. Based on these results, Section~\ref{sec:channel_est} mathematically formulates the discrete-Fresnel domain channel estimation for \ac{OCDM}-based radar systems in detail and presents extensions of the introduced concept to \ac{MU}/\ac{MIMO} and \ac{RadCom} operations. Next, Section~\ref{sec:results} presents a performance analysis of the aforementioned schemes based on simulation and measurement results, and, finally, concluding remarks are given in Section~\ref{sec:conclusion}.

%-----------------------------------------------------------%

\subsection*{Notation}\label{subsec:notation}

Throughout this article, $t$ denotes time in seconds and $f$ denotes frequency in Hertz. Additionally, $\delta[\chi]$, \mbox{$\chi\in\mathbb{Z}$}, is the Kronecker delta function, $\left<\cdot\right>_\upsilon$ is the modulo $\upsilon$ operator, $\upsilon\in\mathbb{N}_+$, and $\mathbb{E}\{\cdot\}$ is the expectation operator.

%***********************************************************%
%-----------------------------------------------------------%
%***********************************************************%

\section{Effects of Time and Frequency Shifts in the Discrete-Fresnel Domain}\label{sec:time-freqConv}

Let the matrix \mbox{$\dot{\mathbf{X}}\in\mathbb{C}^{N\times M}$} be the discrete-Fresnel domain representation of the transmit frame in an \ac{OCDM}-based system, with \mbox{$N\in\mathbb{N}_+$} denoting both the \ac{OCDM} symbol length and the number of subchirps in the \ac{OCDM} system and \mbox{$M\in\mathbb{N}_+$} denoting the total number of \ac{OCDM} symbols in the frame. Before transmission over a channel, $\dot{\mathbf{X}}$ undergoes \acp{IDFnT} along its columns to produce the discrete-time domain representation of the \ac{OCDM} frame \mbox{$\mathbf{x}\in\mathbb{C}^{N\times M}$} \cite{ouyang2016,giroto2020_OCDM_RadCom,giroto2021_tmtt}. Consequently, the relationship between the element \mbox{$x_{n,m}\in\mathbb{C}$} at the $n\mathrm{th}$ row, \mbox{$n\in\{0,1,\dots,N-1\}$}, and $m\mathrm{th}$ column, \mbox{$m\in\{0,1,\dots,M-1\}$}, of $\mathbf{x}$ and the element  \mbox{$\dot{X}_{n,m}\in\mathbb{C}$} at the same position of $\dot{\mathbf{X}}$ is expressed as
\begin{equation}\label{eq:IDFnT}
	x_{n,m} = \frac{1}{N}\e^{\im\frac{\pi}{4}}\sum\limits_{k=0}^{N-1} \dot{X}_{k,m}~\e^{-\im\frac{\pi}{N}(n-k)^2},
\end{equation}
where \mbox{$k\in\{0,1,\dots,N-1\}$} is the subchirp index for all of the $M$ discrete-Fresnel domain representations of \ac{OCDM} symbols contained in $\dot{\mathbf{X}}$. To prevent \ac{ISI}, \acused{CP} cyclic prefixes (CPs) of length $N_\mathrm{CP}$ are prepended to each of the $M$ columns of $\mathbf{x}$, resulting in \mbox{$\mathbf{x}_\mathrm{CP}\in\mathbb{C}^{(N+N_\mathrm{CP})\times M}$}. For successful \ac{ISI} avoidance, it must hold that \mbox{$N_\mathrm{CP}\geq N_\mathrm{h}-1$}, where $N_\mathrm{h}$ is the expected length of the discrete-time domain representation of the \ac{CIR}. Next, $\mathbf{x}_\mathrm{CP}$ undergoes \ac{S/P} conversion, and the real and imaginary parts of the resulting vector individually undergo \ac{D/A} conversion with sampling rate \mbox{$F_\mathrm{s}\geq B$}, where $B$ is the bandwidth occupied by the \ac{OCDM}-based system in Hertz. The output signals by the two aforementioned \acp{DAC} then undergo analog conditioning and I/Q modulation to a carrier with frequency $f_\mathrm{c}\gg B$. Finally, the resulting continuous-time domain transmit signal \mbox{$x(t)\in\mathbb{C}$}, which occupies the \ac{RF} band \mbox{$f\in[f_\mathrm{c}-B/2,f_\mathrm{c}+B/2]$}, is sent out by the transmit antenna of the \ac{OCDM}-based system and propagates through a channel that has \ac{CIR} \mbox{$h(t)\in\mathbb{C}$} and may also introduce frequency shifts in its output signal.

At the receiver side of the \ac{OCDM}-based system, the continuous-time domain output signal from the channel \mbox{$y(t)\in\mathbb{C}$} is received by the receive antenna, undergoing down-conversion and analog conditioning in an I/Q receiver and finally \ac{A/D} conversion with sampling rate $F_\mathrm{s}$. The samples from I and Q channels are then combined into the real and imaginary parts of a serial vector, which after \ac{S/P} conversion becomes the matrix \mbox{$\mathbf{y}_\mathrm{CP}\in\mathbb{C}^{(N+N_\mathrm{CP})\times M}$} that represents the discrete-time domain receive \ac{OCDM} frame. In case \mbox{$F_\mathrm{s}>B$}, it is also assumed that a prior downsampling to $B$ is performed. An inverse processing to the one performed at the transmitter side is then performed on $\mathbf{y}_\mathrm{CP}$, namely, \ac{CP} removal to produce \mbox{$\mathbf{y}\in\mathbb{C}^{N\times M}$}, and column-wise \ac{DFnT}, which ultimately produces the matrix \mbox{$\dot{\mathbf{Y}}\in\mathbb{C}^{N\times M}$} that represents the discrete-Fresnel domain receive \ac{OCDM} frame. The relationship between the element \mbox{$\dot{Y}_{k,m}\in\mathbb{C}$} at the $k\mathrm{th}$ row and $m\mathrm{th}$ column of $\dot{\mathbf{Y}}$ and the element \mbox{$y_{n,m}\in\mathbb{C}$} at the same position of $\mathbf{y}$ is expressed as
\begin{equation}\label{eq:DFnT}
	\dot{Y}_{k,m} = \e^{-\im\frac{\pi}{4}}\sum\limits_{n=0}^{N-1} y_{n,m}~\e^{\im\frac{\pi}{N}(n-k)^2}.
\end{equation}
For the sake of simplicity, \ac{AWGN} is neither considered in \eqref{eq:DFnT} nor in the following expressions throughout this study since the \ac{DFnT} operation does not change its statistics. This is supported by the claims in \cite{ouyang2016} and further discussion on this topic can be found in \cite{omar2021}. %\cite{ouyang2016} below eq (27)

In systems such as wireless communication, radar, and \ac{RadCom} systems, additional processing steps to the aforementioned ones are usually performed. In this section, however, the focus is on analyzing the effects of the propagation through a channel on the originally transmitted \ac{OCDM} symbol representations contained in $\dot{\mathbf{X}}$, i.e., defining the relationship between $\dot{\mathbf{Y}}$ and $\dot{\mathbf{X}}$. The two main effects experienced due to the propagation through a channel are frequency and time shifts. The first are caused by the experienced circular convolution of each of the $M$ columns of $\mathbf{x}$, which represent the discrete-time domain transmit \ac{OCDM} symbols, with the discrete-time domain \ac{CIR} representation \mbox{$\mathbf{h}\in\mathbb{C}^{N_\mathrm{h}\times1}|\mathbf{h}=[h_0, h_1, \dots, h_{N_\mathrm{h}-1}]^T$}. In their turn, the latter effects encompass Doppler shifts caused by reflections off moving scatterers. In this context, Subsections~\ref{subsec:freqShift} and \ref{subsec:timeShift} analyze the sole effects of frequency and time shifts in the discrete-Fresnel domain frame $\dot{\mathbf{Y}}$, respectively, while Subsection~\ref{subsec:freqTimeShift} addresses the case where time and frequency shifts are jointly experienced.

%-----------------------------------------------------------%

\subsection{Effects of frequency shifts in the discrete-Fresnel domain}\label{subsec:freqShift}

If a frequency shift $f_{\Delta}$ is the only experienced effect during propagation of the transmit \ac{OCDM} signal through a channel, then the element at the $n\mathrm{th}$ row and $m\mathrm{th}$ column of the discrete-time domain frame $\mathbf{y}$ obtained after \ac{CP} removal at the receiver side can be expressed as
\begin{eqnarray}\label{eq:yn_Doppler}
	y_{n,m} &=& x_{n,m}~\e^{\im2\pi f_{\Delta}\left[m\left(N+N_\text{CP}\right)+N_\text{CP}+n\right]/B}\nonumber\\
	&=&	x_{n,m}~\e^{\im2\pi f_{\Delta}n/B}\e^{\im2\pi f_{\Delta}\left[m\left(N+N_\text{CP}\right)+N_\text{CP}\right]/B}.
\end{eqnarray}

For the sake of simplicity, the normalized frequency shift $k_{\Delta}\in\mathbb{R}$ and the phase $\phi_m\in\mathbb{R}$ are defined as
\begin{equation}\label{eq:k_delta}
	k_{\Delta} \triangleq f_{\Delta}/(B/N)
\end{equation}
and
\begin{equation}\label{eq:phi_m}
	\phi_m \triangleq 2\pi k_{\Delta}\left[m\left(N+N_\text{CP}\right)+N_\text{CP}\right]/N,
\end{equation}
respectively. Next, performing column-wise \acp{DFnT} on $\mathbf{y}$ yields the matrix $\dot{\mathbf{Y}}$, whose element at the $k\mathrm{th}$ row and $m\mathrm{th}$ column is expressed as
\begin{eqnarray}\label{eq:freqConv1}
	\dot{Y}_{k,m} &=& \e^{-\im\frac{\pi}{4}}\sum\limits_{n=0}^{N-1} \left(x_{n,m}~\e^{\im2\pi k_{\Delta}n/N}\e^{\im\phi_m}\right)~\e^{\im\frac{\pi}{N}(n-k)^2}\nonumber\\
	&=& \e^{\im\phi_m}\e^{-\im\frac{\pi}{4}}\sum\limits_{n=0}^{N-1} x_{n,m}~\e^{\im\frac{\pi}{N}(n^2-2n(k-k_{\Delta})+k^2)}.
\end{eqnarray}
Knowing that
\begin{equation}\label{eq:property1_Doppler}
	n^2 -2nk + k^2 +2nk_{\Delta} = [n-(k-k_{\Delta})]^2 + 2kk_{\Delta} - k_{\Delta}^2,
\end{equation}
it is possible to rewrite \eqref{eq:freqConv1} as
\begin{equation}\label{eq:freqConv2}
	\dot{Y}_{k,m} = \e^{\im\phi_m}\e^{\im\frac{\pi}{N}\left(2kk_{\Delta} - k_{\Delta}^2\right)}\e^{-\im\frac{\pi}{4}}\sum\limits_{n=0}^{N-1} x_{n,m}~\e^{\im\frac{\pi}{N}[n-(k-k_{\Delta})]^2}.
\end{equation}
Expressing $x_{n,m}$ as the \ac{IDFnT} of $\dot{X}_{k,m}$ in \eqref{eq:freqConv2} and rearranging the resulting expression yields
\begin{equation}\label{eq:freqConv3}
	\dot{Y}_{k,m} = \frac{\e^{\im\phi_m}}{N}\sum\limits_{\kappa=0}^{N-1}\dot{X}_{\kappa,m}~\e^{\im\frac{\pi}{N}\left(k^2-\kappa^2\right)}\sum\limits_{n=0}^{N-1}\e^{\im\frac{\pi}{N}n\left(-2k+2k_{\Delta}+2\kappa\right)},
\end{equation}
for \mbox{$\kappa\in\{0,1,\dots,N-1\}$}. The rightmost sum in \eqref{eq:freqConv3} is a finite geometric series, which, according to the result from the Appendix, can be evaluated to yield
\begin{equation}\label{eq:freqConv4}
	\dot{Y}_{k,m} = \frac{\e^{\im\phi_m}}{N}\sum\limits_{\kappa=0}^{N-1}\dot{X}_{\kappa,m}~\e^{\im\frac{\pi}{N}\left(k^2-\kappa^2\right)}\left(\frac{\e^{\im 2\pi\left(k_{\Delta}-k+\kappa\right)}-1}
	{\e^{\im\frac{2\pi}{N}\left(k_{\Delta}-k+\kappa\right)}-1}\right).
\end{equation}
Since $k\in\mathbb{N}$ and $\kappa\in\mathbb{N}$, it holds for the specific case where $k_{\Delta}\in\mathbb{Z}$, which is when the frequency shift $f_{\Delta}$ is an integer multiple of $B/N$, that
\begin{equation}\label{eq:diracFreqConv}
	\left.\frac{\e^{\im 2\pi\left(k_{\Delta}-k+\kappa\right)}-1}
	{\e^{\im\frac{2\pi}{N}\left(k_{\Delta}-k+\kappa\right)}-1}\right|_{k_{\Delta}\in\mathbb{Z}} = \delta\left[\left<\kappa-(k-k_{\Delta})\right>_N\right].
\end{equation}
%\begin{equation}\label{eq:diracFreqConv}
%	\delta\left(\left<\kappa-(k-k_{\Delta})\right>_N\right)
%\end{equation}
The result from \eqref{eq:diracFreqConv} finally allows rewriting \eqref{eq:freqConv4} as
\begin{equation}\label{eq:freqConv5}
	\dot{Y}_{k,m} = \frac{\e^{\im\phi_m}}{N}\sum\limits_{\kappa=0}^{N-1}\dot{X}_{\kappa,m}~\e^{\im\frac{\pi}{N}\left(k^2-\kappa^2\right)}~\delta\left[\left<\kappa-(k-k_{\Delta})\right>_N\right],
\end{equation}
which after further manipulation becomes
%\begin{eqnarray}\label{eq:freqConv6}
%	\dot{Y}_k &=& \dot{X}_{\left<k-k_{\Delta}\right>_N}~\e^{\im\frac{\pi}{N}\left[k^2-\left(k-k_{\Delta}\right)^2\right]}\nonumber\\
%	&=& \dot{X}_{\left<k-k_{\Delta}\right>_N}~\e^{\im\frac{\pi}{N}\left(2kk_{\Delta} - k_{\Delta}^2\right)}
%\end{eqnarray}
\begin{equation}\label{eq:freqConv6}
	\dot{Y}_{k,m} = \e^{\im\phi_m}\dot{X}_{\left<k-k_{\Delta}\right>_N,m}~\e^{\im\frac{\pi}{N}\left(2kk_{\Delta} - k_{\Delta}^2\right)}.
\end{equation}

In summary, the results from \eqref{eq:freqConv4} and \eqref{eq:freqConv6} reveal that a frequency shift $f_{\Delta}$ during the propagation of the transmit \ac{OCDM} signal through a channel results in a circular shift by $k_{\Delta}=f_{\Delta}/B$ samples and a multiplication by the complex exponential \mbox{$\e^{\im\frac{\pi}{N}\left(2kk_{\Delta} - k_{\Delta}^2\right)}$} that rotates the phases of the elements within the columns of $\dot{\mathbf{Y}}$, which represent discrete-Fresnel domain receive \ac{OCDM} symbols. Additionally, the introduced frequency shift rotates the phase of each $m\text{th}$ \ac{OCDM} symbol with respect to the previous one, which is expressed as the phase $\phi_m$ of the complex exponential $\e^{\im\phi_m}$ that multiplies all elements of the $m\mathrm{th}$ column of $\dot{\mathbf{Y}}$.

%\begin{equation}\label{eq:freqConvN}
%	\dot{Y}_k = \e^{\im\frac{\pi}{N}\left[2kk_{\Delta}-k_{\Delta}^2\right]}\dot{X}_{k-k_{\Delta}}
%\end{equation}
%\begin{equation}\label{eq:freqConv3}
%	\dot{Y}_k = \e^{\im\frac{\pi}{N}\left[2k\frac{f_{\Delta}}{\Delta f}-\left(\frac{f_{\Delta}}{\Delta f}\right)^2\right]}\dot{X}_{k-\frac{f_{\Delta}}{\Delta f}}
%\end{equation}

%-----------------------------------------------------------%

\subsection{Effects of time shifts in the discrete-Fresnel domain}\label{subsec:timeShift}

Let an \ac{OCDM} signal be transmitted through a channel with a \ac{CIR} represented in the discrete-time domain by $\mathbf{h}$. In the absence of frequency shifts, the columns of the discrete-time domain receive \ac{OCDM} frame after \ac{CP} removal $\mathbf{y}$ are the result of the circular convolution between the columns of $\mathbf{x}$ and $\mathbf{h}$. In other words, the element at the $n\mathrm{th}$ row and $m\mathrm{th}$ column of $\mathbf{y}$ is expressed as $y_{n,m} = x_{n,m}\circledast h_n$, where $\circledast$ is the circular convolution operator. Alternatively, one can write
\begin{equation}\label{eq:yn_range}
	y_{n,m} = \sum\limits_{\nu=0}^{N-1} h_\nu x_{\left<n-\nu\right>_N,m},
\end{equation}
for \mbox{$\nu\in\{0,1,\dots,N-1\}$}.

By performing column-wise \acp{DFnT} on $\mathbf{y}$, the matrix representation $\dot{\mathbf{Y}}$ of the discrete-Fresnel domain frame is yielded. Its element at the $k\mathrm{th}$ row and $m\mathrm{th}$ column is expressed as
\begin{equation}\label{eq:timeConv1}
	\dot{Y}_{k,m} = \e^{-\im\frac{\pi}{4}}\sum\limits_{n=0}^{N-1}\left(\sum\limits_{\nu=0}^{N-1}h_\nu x_{\left<n-\nu\right>_N,m}\right)~\e^{\im\frac{\pi}{N}(n-k)^2}.
\end{equation}
Substituting $l=\left<n-\nu\right>_N$ in \eqref{eq:timeConv1} and rearranging the resulting expression yields \eqref{eq:timeConv2}. 
\begin{figure*}[!b]
	%\vspace*{4pt}
	\hrulefill
	%\vspace*{1pt}
	\begin{equation}\label{eq:timeConv2}
	\dot{Y}_{k,m} = \sum\limits_{\nu=0}^{N-1}h_\nu~\e^{\im\frac{\pi}{N}(\nu^2-2\nu k)}~\e^{-\im\frac{\pi}{4}}\sum\limits_{l=0}^{N-1}\left(x_{l,m}~\e^{\im 2\pi \nu l/N}\right)~\e^{\im\frac{\pi}{N}(l-k)^2}.
	\end{equation}
\end{figure*}
Based on the obtained results in Subsection~\ref{subsec:freqShift}, more specifically in \eqref{eq:freqConv4}, \eqref{eq:timeConv2} can be rewritten as in \eqref{eq:timeConv3}. 
\begin{figure*}[!b]
	%\vspace*{4pt}
	\hrulefill
	%\vspace*{1pt}
	\begin{equation}\label{eq:timeConv3}
		{Y}_{k,m} = \sum\limits_{\nu=0}^{N-1}h_\nu~\e^{\im\frac{\pi}{N}(\nu^2-2\nu k)}~\left[\frac{1}{N}\sum\limits_{\kappa=0}^{N-1}\dot{X}_{\kappa,m}~\e^{\im\frac{\pi}{N}\left(k^2-\kappa^2\right)}\left(\frac{\e^{\im 2\pi\left(\nu-k+\kappa\right)}-1}
		{\e^{\im\frac{2\pi}{N}\left(\nu-k+\kappa\right)}-1}\right)\right]
	\end{equation}
\end{figure*}
Since $k\in\mathbb{N}$, $\kappa\in\mathbb{N}$, and $\nu\in\mathbb{N}$, the same principle from \eqref{eq:diracFreqConv} can be applied to the rightmost term in \eqref{eq:timeConv3}, which then becomes \eqref{eq:timeConv4}. 
%\begin{equation}\label{eq:diracTimeConv}
%	\frac{\e^{\im 2\pi\left(m-k+\kappa\right)}-1}
%	{\e^{\im\frac{2\pi}{N}\left(m-k+\kappa\right)}-1} = \delta\left(\left<\kappa-(k-m)\right>_N\right)
%\end{equation}
%\begin{equation}\label{eq:diracTimeConv}
%	\delta\left(\left<\kappa-(k-m)\right>_N\right)
%\end{equation}
\begin{figure*}[!b]
	%\vspace*{4pt}
	\hrulefill
	%\vspace*{1pt}
	\begin{equation}\label{eq:timeConv4}
		\dot{Y}_{k,m} = \sum\limits_{\nu=0}^{N-1}h_\nu~\e^{\im\frac{\pi}{N}(\nu^2-2\nu k)}\left(\frac{1}{N}\sum\limits_{\kappa=0}^{N-1}\dot{X}_{\kappa,m}~\e^{\im\frac{\pi}{N}\left(k^2-\kappa^2\right)}~\delta\left(\left<\kappa-(k-\nu)\right>_N\right)\right)
	\end{equation}
\end{figure*}
\begin{figure*}[!b]
	%\vspace*{4pt}
	\hrulefill
	%\vspace*{1pt}
	\setcounter{equation}{22}
	\begin{eqnarray}\label{eq:timeFreqConv3}
	\dot{Y}_{k,m} = \frac{\e^{\im\phi_m}}{N}\sum\limits_{\kappa=0}^{N-1}\left(\sum\limits_{\nu=0}^{N-1}h_\nu~\dot{X}_{\left<\kappa-\nu\right>_N,m}\right)~\e^{\im\frac{\pi}{N}\left(k^2-\kappa^2\right)}~\left(\frac{\e^{\im 2\pi\left(k_{\Delta}-k+\kappa\right)}-1}
	{\e^{\im\frac{2\pi}{N}\left(k_{\Delta}-k+\kappa\right)}-1}\right)%~\delta\left(\left<\kappa-(k-k_{\Delta})\right>_N\right)
	%\dot{Y}_k = \frac{1}{N}\sum\limits_{\kappa=0}^{N-1}\left[\sum\limits_{m=0}^{N-1}h_m~\e^{\im\frac{\pi}{N}(m^2-2m\kappa)}\left(\frac{1}{N}\sum\limits_{\kappa'=0}^{N-1}\dot{X}_{\kappa'}~\e^{\im\frac{\pi}{N}\left(\kappa^2-\kappa'^2\right)}~\delta[\left<\kappa'-(\kappa-m)\right>_N]\right)\right]~\e^{\im\frac{\pi}{N}\left(k^2-\kappa^2\right)}~\delta[\left<\kappa-(k-k_{\Delta})\right>_N]
	\end{eqnarray}
\end{figure*}
After further manipulation, \eqref{eq:timeConv4} can be finally rewritten as
\setcounter{equation}{18}
\begin{equation}\label{eq:timeConv5}
	\dot{Y}_{k,m} = \sum\limits_{\nu=0}^{N-1}h_\nu\dot{X}_{\left<k-\nu\right>_N,m}
\end{equation}

The relationship between $\dot{Y}_{k,m}$ and $\dot{X}_{k,m}$ as in \eqref{eq:timeConv5} reveals that the same circular shifts and amplitude weightings suffered by the columns of $\mathbf{x}$ are observed in the columns of both $\mathbf{y}$  and its discrete-Fresnel domain representation $\dot{\mathbf{Y}}$. In other words, \mbox{$y_{n,m} = x_{n,m}\circledast h_n$} results in \mbox{$\dot{Y}_{k,m} = \dot{X}_{k,m}\circledast h_k$}, which is also known as the convolution theorem of the \ac{DFnT} \cite{ouyang2015,ouyang2016}.

%-----------------------------------------------------------%

\subsection{Joint effects of time and frequency shifts in the discrete-Fresnel domain}\label{subsec:freqTimeShift}

Should the propagation of the \ac{OCDM} transmit signal through a channel results not only in time shifts represented by the \ac{CIR} $\mathbf{h}$, but also in a normalized frequency shift $k_{\Delta}$, the element at the $n\mathrm{th}$ row and $m\mathrm{th}$ column of the discrete-time domain receive frame $\mathbf{y}$ obtained after \ac{CP} removal at the receiver side can be expressed as
\begin{equation}\label{eq:yn_rangeDoppler}
	y_{n,m} = \left(x_{n,m}\circledast h_n\right)~\e^{\im2\pi k_{\Delta}n/N}\e^{\im\phi_m}.
\end{equation}
For the sake of simplicity, the auxiliary matrix \mbox{$\mathbf{r}\in\mathbb{C}^{N\times M}$} is defined, being the element \mbox{$r_{n,m}\in\mathbb{C}$} located at its $n\mathrm{th}$ row and $m\mathrm{th}$ column given by
\begin{equation}\label{eq:rn_conv}
	r_{n,m} \triangleq x_{n,m}\circledast h_n.
\end{equation}
Consequently, \eqref{eq:yn_rangeDoppler} can be alternatively expressed as
\begin{equation}\label{eq:rn_Doppler}
	y_{n,m} = r_{n,m}~\e^{\im2\pi k_{\Delta}n/N}\e^{\im\phi_m}.
\end{equation}
Defining \mbox{$\dot{\mathbf{R}}\in\mathbb{C}^{N\times M}$} as the output of column-wise \acp{DFnT} on $\mathbf{r}$, the relationship between the element \mbox{$\dot{R}_{k,m}\in\mathbb{C}$} at the $k\mathrm{th}$ row and $m\mathrm{th}$ column of  $\dot{\mathbf{R}}$ and the element $\dot{X}_{k,m}$ at the same position of \mbox{$\dot{\mathbf{X}}$} can be defined based on the result from \eqref{eq:timeConv4}. Additionally, the element $\dot{Y}_{k,m}$ at the $k\mathrm{th}$ row and $m\mathrm{th}$ column of \mbox{$\dot{\mathbf{Y}}$} can be derived from $\dot{R}_{k,m}$ based on the result from \eqref{eq:freqConv4}. Consequently, $\dot{Y}_{k,m}$ can be expressed as a function of $\dot{X}_{k,m}$ as in \eqref{eq:timeFreqConv3}.
In the specific case where $k_{\Delta}\in\mathbb{Z}$, the property used in \eqref{eq:diracFreqConv} can be finally applied to \eqref{eq:timeFreqConv3}, yielding
\setcounter{equation}{23}
\begin{equation}\label{eq:timeFreqConv4}
	\dot{Y}_{k,m} = \e^{\im\phi_m}\left(\sum\limits_{\nu=0}^{N-1}h_\nu\dot{X}_{\left<(k-k_{\Delta})-\nu\right>_N,m}\right)~\e^{\im\frac{\pi}{N}\left(2kk_{\Delta}-k_{\Delta}^2\right)},
\end{equation}
i.e., $\dot{Y}_{k,m} = \e^{\im\phi_m}(\dot{X}_{\left<k-k_{\Delta}\right>_N,m}\circledast h_k)~\e^{\im\frac{\pi}{N}\left(2kk_{\Delta}-k_{\Delta}^2\right)}$.

The results from \eqref{eq:timeFreqConv3} and \eqref{eq:timeFreqConv4} indicate that the convolution with the \ac{CIR} followed by a frequency shift $f_{\Delta}$ during the propagation of the transmit \ac{OCDM} signal through a channel results in a circular shift by $k_{\Delta}=f_{\Delta}/B$ samples on top of the circular shifts caused by the convolution with the multiple taps of the discrete-time domain \ac{CIR} representation $\mathbf{h}$. This effect is similar to known range- or delay-Doppler coupling in chirp-based radar systems \cite{winkler2007}. Besides the aforementioned effect, a multiplication by the complex exponential \mbox{$\e^{\im\frac{\pi}{N}\left(2kk_{\Delta} - k_{\Delta}^2\right)}$} of the shifted versions of $\dot{\mathbf{X}}$ takes place as observed in Subsection~\ref{subsec:freqShift}, increasingly rotating the phases of the $N$ elements of each column of $\dot{\mathbf{Y}}$. As in the case of Subsection~\ref{subsec:freqShift}, every $m\mathrm{th}$ column of $\dot{\mathbf{Y}}$ is additionally multiplied by the complex exponential $\e^{\im\phi_m}$, which rotates the phase of all elements of the $m\mathrm{th}$ discrete-Fresnel domain \ac{OCDM} symbol. The aforementioned effects ultimately distort the transmit subchirps and cause \ac{IChI} in the discrete-Fresnel domain, which is somewhat similar to the \ac{ICI} in the discrete-frequency domain in \ac{OFDM}-based systems.

%\begin{equation}\label{eq:timeFreqConv}
%	\dot{Y}_k = \sum\limits_{h=0}^{H-1}\alpha_h\e^{\im\frac{\pi}{N}\left[2k\frac{f_{\mathrm{D},h}}{\Delta f}-\left(\frac{f_{\mathrm{D},h}}{\Delta f}\right)^2\right]}\dot{X}_{k-\frac{2BR_h}{c_0}-\frac{f_{\mathrm{D},h}}{\Delta f}}
%\end{equation}

%***********************************************************%
%-----------------------------------------------------------%
%***********************************************************%

\section{Discrete-Fresnel Domain Channel Estimation in OCDM-based Radar Systems}\label{sec:channel_est}

\begin{figure*}[!b]
	\setcounter{equation}{25}
	%\vspace*{4pt}
	\hrulefill
	%\vspace*{1pt}
	\begin{equation}\label{eq:yRad_ideal}
	\tilde{y}^\mathrm{rad}_{n,m}
	= \sum\limits_{\eta=0}^{H-1}\left[\sum\limits_{\nu=0}^{N-1}x_{\nu,m}~\left(\frac{\e^{\im 2\pi\left(n_{\Delta,\eta}-n+\nu\right)}-1}
	{\e^{\im\frac{2\pi}{N}\left(n_{\Delta,\eta}-n+\nu\right)}-1}\right)\right]~\e^{\im 2\pi k_{\Delta,\eta}n/N}\e^{\im \phi_{m,\eta}}
	%&=& \sum\limits_{\eta=0}^{H-1}\left(\sum\limits_{\nu=0}^{N-1}x_{\nu,m}h^{\mathrm{rad},\eta}_{\left<n-\nu\right>_N}\right)~\e^{\im 2\pi k_{\Delta,\eta}n/N}\e^{\im \phi_{m,\eta}}\nonumber\\
	\end{equation}
\end{figure*}

The discussion in Section~\ref{sec:time-freqConv} described the effects of time and frequency shifts on the originally transmitted discrete-Fresnel domain \ac{OCDM} symbols represented by the columns of $\dot{\mathbf{X}}$, which becomes $\dot{\mathbf{Y}}$ after considering the effects of the propagation through a channel. Based on the achieved results, it is possible to predict the relationship between the channel estimates in \ac{OCDM}-based systems and the actual channel \ac{CIR}.

For this purpose, the discrete-Fresnel domain pilot design strategy introduced in \cite{ouyang2017_unbiasedCE} for coherent optical \ac{OFDM} systems, which was also investigated for reflectometric sensing of power lines in baseband \ac{OCDM} systems as reported in \cite{giroto2021} and further improved in \cite{ouyang2018_unbiasedCE}, is considered. The original pilot design from \cite{ouyang2017_unbiasedCE} performs unbiased channel estimation under the assumption that no frequency shifts take place. This is done by exploiting the convolution theorem of the \ac{DFnT} \cite{ouyang2015} and having a Kronecker comb with \mbox{$P\in\mathbb{N}_+$} active subchirps in the discrete-Fresnel domain at the transmitter side, and averaging the resulting $P$ sections of length $N/P$ in the discrete-Fresnel domain at the receiver side to estimate the \ac{CIR}. Conversely, the improvement proposed in \cite{ouyang2018_unbiasedCE} achieves optimal \ac{CIR} estimates in the \ac{MMSE} sense by activating only the first subchirp in the discrete-Fresnel domain at the transmitter side, and discarding noisy samples beyond the maximum expected \ac{CIR} length $N_\mathrm{h}$ in the obtained discrete-Fresnel domain receive \ac{OCDM} symbol to obtain a \ac{CIR} estimate. Since dividing the transmission power among $P$ active subchirps and averaging their corresponding \ac{CIR} estimates yields the same \ac{SNR} as in the case where the full transmission power is allocated to a single subchirp and only one \ac{CIR} is obtained, the superiority of the improved pilot design approach is solely due to the noise-rejection windowing performed in the discrete-Fresnel domain at the receiver side. 

Considering the two aforementioned characteristics of the pilot design strategies, namely the one from \cite{ouyang2017_unbiasedCE} and the one from \cite{ouyang2018_unbiasedCE}, the latter one is adopted in this study. %\cor{For communication and range-only radar sensing purposes, one can strategically reserve \ac{OCDM} symbols within a frame depending on expected \ac{CFO} and/or Doppler shifts for either transmitting a single pilot \ac{OCDM} symbol or multiple ones for coherent accumulation aiming for \ac{SNR} improvement. In radar-only or \ac{ISAC} applications with simultaneous range and velocity sensing, however, one would consecutively transmit multiple identical pilot \ac{OCDM} symbols, which could comprise the whole \ac{OCDM} frame or a preamble as in the \ac{TDD} between signals for \ac{CS}-based radar sensing and communication described in \cite{pham2020,dwivedi2020,barreto2021}. More efficient strategies for \ac{ISAC} would, for instance, include the design of discrete-Fresnel domain transmit \ac{OCDM} symbols for jointly transmitting data and performing channel estimation. However, as channel estimation in the discrete-Fresnel domain under the simultaneous effect of time and frequency shifts has not been yet addressed in the literature, this study will solely consider pilot \ac{OCDM} symbols as described in \cite{ouyang2018_unbiasedCE}.}
Consequently, the element at the $k\mathrm{th}$ row and $m\mathrm{th}$ column of the matrix $\mathbf{\dot{X}}$ that represents a frame containing $M$ discrete-Fresnel domain pilot \ac{OCDM} symbols in its columns is given by
\setcounter{equation}{24}
\begin{equation}\label{eq:FresnelPilot}
	\dot{X}_{k,m} = \delta[k].
\end{equation}
As all $M$ \ac{OCDM} symbols within the frame are equal, no \ac{CP} is required. 
%The implications of the pilot \ac{OCDM} symbol design on channel estimation for radar and communication purposes are discussed in Subsections~\ref{subsec:radarChannelEst} and \ref{subsec:commChannelEst}, respectively.
The implications of the pilot \ac{OCDM} symbol design on the performance of radar channel estimation is analyzed in Subsection~\ref{subsec:radarChannelEst}. Furthermore, extensions of the discussed concept to \ac{MU} or \ac{MIMO} scenarios as well as to \ac{RadCom} applications are discussed in Subsections~\ref{subsec:multiplexing} and \ref{subsec:radCom}, respectively.
%
%\cor{----------}
%
%\cor{Extension to multiple OCDM symbols and correction of the radar channel estimation formulation (each target with its own delay and Doppler shift) after this point is still pending.}
%
%\cor{----------}

%-----------------------------------------------------------%

\subsection{Radar Channel Estimation}\label{subsec:radarChannelEst}

For the sake of simplicity, it is assumed in this subsection that the \ac{OCDM}-based system acts as a monostatic radar, in which transmit and receive antennas are virtually collocated. In practice, however, only a quasi-monostatic radar is possible, in which the distance between transmit and receive antennas is negligible in comparison to the range of expected target distances w.r.t. the radar. To estimate the range and relative radial velocity of multiple point targets w.r.t. the radar, consecutive radar \acp{CIR} are estimated. Each \ac{CIR} is composed of multiple taps at delays corresponding to the \ac{ToF} taken by the \ac{OCDM} signal to be sent out by the transmit antenna, reflected off targets, and captured by the receive antenna. If no effects such as range migration occur within the measurement time, all of the estimated radar \acp{CIR} allow estimating the range of resolved point targets present in the monitored scenario. The observed difference among consecutive \acp{CIR}, which is the phase progression of its taps, ultimately allows obtaining a radar image that has not only range information, but also estimates of the experienced Doppler shifts related to the relative radial velocities of the point targets. To enable the aforementioned estimation of range and Doppler shifts of \mbox{$H\in\mathbb{N}$} point targets, the matrix $\tilde{\mathbf{y}}^\mathrm{rad}\in\mathbb{C}^{N\times M}$ that would ideally represent the discrete-time domain receive \ac{OCDM} frame has elements at its $n\mathrm{th}$ row and $m\mathrm{th}$ column expressed as in \eqref{eq:yRad_ideal}. In this equation, \mbox{$n_{\Delta,\eta}\in\mathbb{R}|n_{\Delta,\eta}=2R_\eta B/c_0$} is a normalized range term, being \mbox{$R_\eta\in\mathbb{R}_+$} the range in meters of the $\eta\mathrm{th}$ target and $c_0$ the speed of light in vacuum in meters per second. Additionally, \mbox{$k_{\Delta,\eta}\in\mathbb{R}$} and \mbox{$\phi_{m,\eta}\in\mathbb{R}$} are the normalized Doppler shift and the Doppler phase associated to the $\eta\mathrm{th}$ point target. In the specific case where \mbox{$n_{\Delta,\eta}\in\mathbb{Z}$}, \eqref{eq:yRad_ideal} can be rewritten as
\setcounter{equation}{26}
\begin{equation}\label{eq:yRad_ideal_int}
	\tilde{y}^\mathrm{rad}_{n,m} = \sum\limits_{\eta=0}^{H-1}x_{n,m}~\delta[n-n_{\Delta,\eta}]~\e^{\im 2\pi k_{\Delta,\eta}n/N}\e^{\im \phi_{m,\eta}}.
\end{equation}

With the \ac{OCDM} pilot symbols contained in the columns of $\dot{\mathbf{X}}$ having their $k\mathrm{th}$ elements defined as in \eqref{eq:FresnelPilot}, the resulting ideal discrete-Fresnel domain receive \ac{OCDM} frame matrix representation $\tilde{\dot{\mathbf{Y}}}^\mathrm{rad}$ can be alternatively defined as the matrix \mbox{$\tilde{\mathbf{h}}^\mathrm{rad}\in\mathbb{C}^{N\times M}$} that contains $M$ consecutive radar \ac{CIR} estimates of length $N$, which can be expressed as
\begin{equation}
	\tilde{h}^\mathrm{rad}_{n,m} = \sum\limits_{\eta=0}^{H-1}\left(\frac{\e^{\im 2\pi\left(n_{\Delta,\eta}-n\right)}-1}
	{\e^{\im\frac{2\pi}{N}\left(n_{\Delta,\eta}-n\right)}-1}\right)~\e^{\im 2\pi k_{\Delta,\eta}n/N}\e^{\im \phi_{m,\eta}}
\end{equation}
or simply
\begin{equation}
	\tilde{h}^\mathrm{rad}_{n,m} = \sum\limits_{\eta=0}^{H-1}\delta[n-n_{\Delta,\eta}]~\e^{\im 2\pi k_{\Delta,\eta}n/N}\e^{\im \phi_{m,\eta}}
\end{equation}
for \mbox{$n_{\Delta,\eta}\in\mathbb{Z}$}. 

\begin{figure}[!t]
	\centering
	\psfrag{A}[c][c]{\scriptsize $f_\mathrm{c}-B/2$}
	\psfrag{B}[c][c]{\scriptsize $f_\mathrm{c}$}
	\psfrag{C}[c][c]{\scriptsize $f_\mathrm{c}+B/2$}
	\psfrag{D}[c][c]{\scriptsize $0$}
	\psfrag{E}[c][c]{\scriptsize $N/2$}
	\psfrag{F}[c][c]{\scriptsize $N-1$}
	\psfrag{G}[c][c]{\footnotesize $\angle \hat{H}^\mathrm{rad}_m(f)$}
	\psfrag{J}[c][c]{\footnotesize $\angle \hat{H}^\mathrm{rad}_{k,m}$}
	\psfrag{H}[c][c]{\footnotesize $f$}
	\psfrag{I}[c][c]{\footnotesize $k$}
	\includegraphics[width=8.5cm]{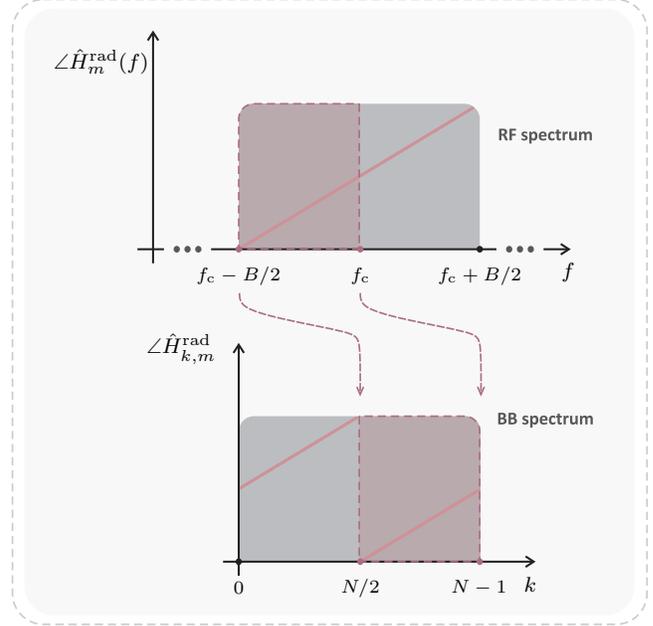}
	\caption{Channel frequency response phase progression due to experienced propagation time delay in RF and BB spectra \cite{nuss2021_diss}.}\label{fig:phase}
	%\vspace{-0.25cm}	
\end{figure}

In reality, however, a matrix \mbox{$\dot{\mathbf{Y}}^\mathrm{rad}\in\mathbb{C}^{N\times M}|\dot{\mathbf{Y}}^\mathrm{rad}\neq \tilde{\dot{\mathbf{Y}}}^\mathrm{rad}$} is obtained at the receiver side of the \ac{OCDM}-based system. Since pilot \ac{OCDM} symbols are transmitted, $\dot{\mathbf{Y}}^\mathrm{rad}$ can be alternatively defined as the matrix \mbox{$\hat{\mathbf{h}}^\mathrm{rad}\in\mathbb{C}^{N\times M}|\hat{\mathbf{h}}^\mathrm{rad}\neq \tilde{\mathbf{h}}^\mathrm{rad}$} that contains the actual consecutive radar \ac{CIR} estimates in its columns. Assuming a single target scenario and considering only the $m\mathrm{th}$ of the $M$ radar channel estimates, a corresponding continuous-frequency domain \ac{CFR} \mbox{$\hat{H}^\mathrm{rad}_m(f)\in\mathbb{C}$} of the aforementioned channel presents linear phase progression along with the frequency in the \ac{RF} band \mbox{$f\in[f_\mathrm{c}-B/2,f_\mathrm{c}+B/2]$} as represented at the top of Fig.~\ref{fig:phase}. In the discrete-frequency domain, the \ac{CIR} contained in the $m\mathrm{th}$ column of $\hat{\mathbf{h}}^\mathrm{rad}$ is represented by the vector \mbox{$\hat{\mathbf{H}}^\mathrm{rad}_m\in\mathbb{C}^{N\times 1}$}. Since the samples \mbox{$k=0$} to \mbox{$k=N/2-1$} of $\hat{\mathbf{H}}^\mathrm{rad}_m$ can be used to reconstruct spectral information on the \ac{RF} frequencies $f_\mathrm{c}$ to \mbox{$f_\mathrm{c}+(N/2-1)B/N = f_\mathrm{c}+B/2-B/N$}, while the samples \mbox{$k=N/2+1$} to \mbox{$k=N$} allow reconstructing spectral information on the frequencies \mbox{$f_\mathrm{c}+(-N/2)B/N = f_\mathrm{c}-B/2$} to \mbox{$f_\mathrm{c}-B/N$}, the phase of  $\hat{\mathbf{H}}^\mathrm{rad}_m$ is circularly shifted by $N/2$ samples w.r.t. an ideally linearly increasing phase \cite{nuss2021_diss}. Considering a single-target scenario, an exemplary phase progression in the \ac{BB} of $\hat{\mathbf{H}}^\mathrm{rad}_m$ can be seen in Fig.~\ref{fig:phase}.%

In order for the \acp{ToF}, and consequently target ranges, to be appropriately estimated, the effectively obtained matrix \mbox{$\mathbf{y}^\mathrm{rad}\in\mathbb{C}^{N\times M}|\mathbf{y}^\mathrm{rad}\neq\tilde{\mathbf{y}}^\mathrm{rad}$} that represents the discrete-time domain receive \ac{OCDM} frame must consist of circularly-shifted versions of $\mathbf{x}$. If the experienced phase folding is not compensated, the columns of the estimated radar \ac{CIR} matrix $\hat{\mathbf{h}}^\mathrm{rad}$ are equal to the corresponding columns of the ideally obtained matrix $\tilde{\mathbf{h}}^\mathrm{rad}$ with the addition of discrete-frequency domain shift by $N/2$ samples. In other words
\begin{equation}
	\hat{h}^\mathrm{rad}_{n,m} = \tilde{h}^\mathrm{rad}_{n,m}\e^{-\im 2\pi(N/2)n/N} = \tilde{h}^\mathrm{rad}_{n,m}\e^{-\im\pi n},
\end{equation}
where $\hat{h}^\mathrm{rad}_{n,m}$ denotes the element of $\hat{\mathbf{h}}^\mathrm{rad}$ at its $n\mathrm{th}$ row and $m\mathrm{th}$ column.
Although the amplitudes of the elements of $\hat{\mathbf{h}}^\mathrm{rad}$ are the same as the corresponding ones of $\tilde{\mathbf{h}}^\mathrm{rad}$, the introduced phase rotations would ultimately prevent correctly estimating the ranges of targets when reconstructing the analog equivalent of $\hat{\mathbf{h}}^\mathrm{rad}$, e.g., via \ac{ZP}. Consequently, the experienced phase folding must be corrected by performing a circular shift in the discrete-frequency domain by $N/2$ on $\dot{\mathbf{Y}}^\mathrm{rad}$ to produce \mbox{$\dot{\mathbf{Y}}^\mathrm{rad,corr}\in\mathbb{C}^{N\times M}$}, which can be alternatively defined as the matrix \mbox{$\hat{\mathbf{h}}^\mathrm{rad,corr}\in\mathbb{C}^{N\times M}$} that contains an estimate of the desired radar \ac{CIR} matrix $\tilde{\mathbf{h}}^\mathrm{rad}$. The aforementioned phase folding compensation can be achieved in two manners. The first one is by multiplying every $k\mathrm{th}$ element of 
$\dot{\mathbf{Y}}^\mathrm{rad}$ by $\e^{\im\pi k}$. A second solution is based on the \ac{OCDM} receiver \#2 structure from \cite{ouyang2016}, where the \ac{DFnT} is calculated by means of a \ac{DFT}, followed by an element-wise multiplication by the vector \mbox{$\mathbf{\Gamma}\in\mathbb{C}^{N\times 1}|\mathbf{\Gamma}=[\Gamma_0, \Gamma_1, \dots, \Gamma_{N-1}]^T$}, whose $k\text{th}$ element is expressed for even $N$ as
\begin{equation}\label{eq:OCDM_Gamma}
	\Gamma_k = \e^{-\im \pi k^2/N},
\end{equation}
and an \ac{IDFT}. In this approach, a circular shift in the discrete-frequency domain by $N/2$ samples is performed between the element-wise multiplication by $\mathbf{\Gamma}$ and the \ac{IDFT}.

Assuming that no Doppler shifts take place, i.e., \mbox{$k_{\Delta,\eta}=0$} and \mbox{$\phi_{m,\eta}=0$} $\forall m,\eta$, both described phase folding correction approaches yield the \ac{CIR} matrix estimate $\hat{\mathbf{h}}^\mathrm{rad,corr}$, whose element at its $n\mathrm{th}$ row and $m\mathrm{th}$ column is given by
\begin{equation}\label{eq:timeConv_rad2}
	\hat{h}^\mathrm{rad,corr}_{n,m} = \e^{\im\pi n}\hat{h}^\mathrm{rad}_{n,m} = \tilde{h}^\mathrm{rad}_{n,m}.
%	\dot{Y}^\mathrm{rad,corr}_{k,m} &=& \e^{\im\pi k}\dot{Y}_k\nonumber\\
%	&=& \e^{\im\pi k}\left(\tilde{h}^\mathrm{rad}_{k,m}~\e^{-\im\pi k}\right)\nonumber\\
%	&=& \tilde{h}^\mathrm{rad}_{k,m}	
\end{equation}
If, however, a frequency shift also takes place, the discrete-time domain \ac{CIR} matrix representation without phase folding correction $\hat{\mathbf{h}}^\mathrm{rad}$ has the element at its $n\mathrm{th}$ row and $m\mathrm{th}$ column expressed as in \eqref{eq:timeFreqConv_rad1} or, for the specific case where $n_{\Delta,\eta}\in\mathbb{Z}$ and $k_{\Delta,\eta}\in\mathbb{Z}$ $\forall \eta$, as in \eqref{eq:timeFreqConv_rad2}.
\begin{figure*}[!t]
	%\vspace*{4pt}
	%\hrulefill
	%\vspace*{1pt}
	%\setcounter{equation}{27}
	\begin{eqnarray}\label{eq:timeFreqConv_rad1}
	\hat{h}^\mathrm{rad}_{n,m} &=& \sum\limits_{\eta=0}^{H-1}\frac{\e^{\im \phi_{m,\eta}}}{N}\sum\limits_{\kappa=0}^{N-1}\left[\sum\limits_{\nu=0}^{N-1}\left(\frac{\e^{\im 2\pi\left(n_{\Delta,\eta}-\nu\right)}-1}
	{\e^{\im\frac{2\pi}{N}\left(n_{\Delta,\eta}-\nu\right)}-1}\right)~\e^{-\im\pi \nu}~\delta[\left<\kappa-\nu\right>_N]\right]~\e^{\im\frac{\pi}{N}\left(n^2-\kappa^2\right)}~\left(\frac{\e^{\im 2\pi\left(k_{\Delta,\eta}-n+\kappa\right)}-1}
	{\e^{\im\frac{2\pi}{N}\left(k_{\Delta,\eta}-n+\kappa\right)}-1}\right)\nonumber\\%~\delta[\left<\kappa-(n-k_{\Delta})\right>_N]
	%%%%%%%%%%%%%%%%%%%%%%%%%%%%%%%%%%%%%%%%%%%%%%%%%%%%%%%%%%%
	&=& \sum\limits_{\eta=0}^{H-1}\frac{\e^{\im \phi_{m,\eta}}}{N}\sum\limits_{\kappa=0}^{N-1}\left[\left(\frac{\e^{\im 2\pi\left(n_{\Delta,\eta}-\kappa\right)}-1}
	{\e^{\im\frac{2\pi}{N}\left(n_{\Delta,\eta}-\kappa\right)}-1}\right)~\e^{-\im\pi \kappa}\right]~\e^{\im\frac{\pi}{N}\left(n^2-\kappa^2\right)}~\left(\frac{\e^{\im 2\pi\left(k_{\Delta,\eta}-n+\kappa\right)}-1}
	{\e^{\im\frac{2\pi}{N}\left(k_{\Delta,\eta}-n+\kappa\right)}-1}\right)%~\delta[\left<\kappa-(n-k_{\Delta})\right>_N]
	\end{eqnarray}
\end{figure*}
\begin{figure*}[!t]
	%\vspace*{4pt}
	%\hrulefill
	%\vspace*{1pt}
	%\setcounter{equation}{27}
	\begin{eqnarray}\label{eq:timeFreqConv_rad2}
		\hat{h}^\mathrm{rad}_{n,m} &=& \sum\limits_{\eta=0}^{H-1}\e^{\im \phi_{m,\eta}}\delta[\left<n-(n_{\Delta,\eta}+k_{\Delta,\eta})\right>_N]~\e^{-\im\pi[n-(n_{\Delta,\eta}+k_{\Delta,\eta})]}~\e^{\im\frac{\pi}{N}\left(n^2-(n-k_{\Delta,\eta})^2\right)}\nonumber\\
		%%%%%%%%%%%%%%%%%%%%%%%%%%%%%%%%%%%%%%%%%%%%%%%%%%%%%%%%%%%
		&=& \sum\limits_{\eta=0}^{H-1}\e^{\im \phi_{m,\eta}}\delta[\left<n-(n_{\Delta,\eta}+k_{\Delta,\eta})\right>_N]~\e^{-\im\pi[n-(n_{\Delta,\eta}+k_{\Delta,\eta})]}~\e^{\im\frac{\pi}{N}\left(2nk_{\Delta,\eta}-k_{\Delta,\eta}^2\right)}
	\end{eqnarray}
	\hrulefill
\end{figure*}
After performing phase folding correction, $\hat{\mathbf{h}}^\mathrm{rad,corr}$ is obtained such that the element in its $n\mathrm{th}$ row and $m\mathrm{th}$ column is expressed as in \eqref{eq:timeFreqConv_rad3} or, for $n_{\Delta,\eta}\in\mathbb{Z}$ and $k_{\Delta,\eta}\in\mathbb{Z}$ $\forall \eta$, as in \eqref{eq:timeFreqConv_rad4}.

\begin{figure*}[!t]
	%\vspace*{4pt}
	%\hrulefill
	%\vspace*{1pt}
	%\setcounter{equation}{27}
	\begin{equation}\label{eq:timeFreqConv_rad3}
	\hat{h}^\mathrm{rad,corr}_{n,m} = \e^{\im\pi n}\hat{h}^\mathrm{rad}_{n,m}
	%%%%%%%%%%%%%%%%%%%%%%%%%%%%%%%%%%%%%%%%%%%%%%%%%%%%%%%%%%%
	 %&=& \e^{\im\pi n}\left\{\frac{1}{N}\sum\limits_{\kappa=0}^{N-1}\left[\sum\limits_{m=0}^{N-1}\left(h^\mathrm{rad}_m~\e^{-\im\pi m}\right)~\delta[\left<\kappa-m\right>_N]\right]~\e^{\im\frac{\pi}{N}\left(n^2-\kappa^2\right)}~\left(\frac{\e^{\im 2\pi\left(k_{\Delta}-n+\kappa\right)}-1}
	%{\e^{\im\frac{2\pi}{N}\left(k_{\Delta}-n+\kappa\right)}-1}\right)\right\}\nonumber\\%~\delta[\left<\kappa-(n-k_{\Delta})\right>_N]
	%%%%%%%%%%%%%%%%%%%%%%%%%%%%%%%%%%%%%%%%%%%%%%%%%%%%%%%%%%%
	= \e^{\im\pi n}\left\{\sum\limits_{\eta=0}^{H-1}\frac{\e^{\im \phi_{m,\eta}}}{N}\sum\limits_{\kappa=0}^{N-1}\left[\left(\frac{\e^{\im 2\pi\left(n_{\Delta,\eta}-\kappa\right)}-1}
	{\e^{\im\frac{2\pi}{N}\left(n_{\Delta,\eta}-\kappa\right)}-1}\right)~\e^{-\im\pi \kappa}\right]~\e^{\im\frac{\pi}{N}\left(n^2-\kappa^2\right)}~\left(\frac{\e^{\im 2\pi\left(k_{\Delta,\eta}-n+\kappa\right)}-1}
	{\e^{\im\frac{2\pi}{N}\left(k_{\Delta,\eta}-n+\kappa\right)}-1}\right)\right\}%~\delta[\left<\kappa-(n-k_{\Delta})\right>_N]
	\end{equation}
\end{figure*}
\begin{figure*}[!t]
	%\vspace*{4pt}
	%\hrulefill
	%\vspace*{1pt}
	%\setcounter{equation}{27}
	\begin{equation}\label{eq:timeFreqConv_rad4}
		\hat{h}^\mathrm{rad,corr}_{n,m}
		= \e^{\im\pi n}\hat{h}^\mathrm{rad}_{n,m}
		= \sum\limits_{\eta=0}^{H-1}\e^{\im \phi_{m,\eta}}\delta[\left<n-(n_{\Delta,\eta}+k_{\Delta,\eta})\right>_N]~\e^{\im\frac{\pi}{N}\left[2nk_{\Delta,\eta}-k_{\Delta,\eta}^2+N(n_{\Delta,\eta}+k_{\Delta,\eta})\right]}
	\end{equation}
	\hrulefill
\end{figure*}

The obtained expressions in \eqref{eq:timeFreqConv_rad3} and \eqref{eq:timeFreqConv_rad4} reveal that the delay-Doppler coupling mentioned in Subsection~\ref{subsec:freqTimeShift} still takes place after phase folding correction. Additionally, it is observed for the $\eta\mathrm{th}$ target contribution to the aforementioned equations that the columns of $\hat{\mathbf{h}}^\mathrm{rad,corr}$ are circularly-shifted versions of the columns of $\tilde{\mathbf{h}}^\mathrm{rad}$ by $k_{\Delta,\eta}$ samples. Compared to their counterparts in $\tilde{\mathbf{h}}^\mathrm{rad}$, the $n\mathrm{th}$ elements of the aforementioned columns are also multiplied by complex exponentials with linearly increasing phase along with the index $n$.
%
%\corb{
%No Doppler shift for simplicity \cor{(maybe add Doppler to delete previous equations)}: ideal channel
%\begin{eqnarray}
%	Y^\mathrm{rad}_{l,m} &=& \sum\limits_{\eta=0}^{H-1}X_{l,m}~H^{\mathrm{rad},\eta}_l\nonumber\\
%	&=& \sum\limits_{\eta=0}^{H-1}X_{l,m}~\e^{\im 2\pi n_{\Delta,\eta} l/N}	
%\end{eqnarray}
%}
%
%\corb{
%	No Doppler shift for simplicity: real channel
%	\begin{eqnarray}
%	Y^\mathrm{rad}_{l,m} &=& \sum\limits_{\eta=0}^{H-1}X_{l,m}~H^\eta_l\nonumber\\
%	&=& \sum\limits_{\eta=0}^{H-1}X_{l,m}~\e^{\im 2\pi n_{\Delta,\eta} \left<l\right>_{N/2}/N}	
%	\end{eqnarray}
%}
%
%\corb{
%\begin{equation}
%	h_n = h^\mathrm{rad}_n\e^{-\im 2\pi(N/2)n/N} = h^\mathrm{rad}_n\e^{-\im\pi n}
%\end{equation}
%}
%
%\corb{
%	\begin{equation}
%	H_l = H^\mathrm{rad}_{\left<l\right>_{N/2}}
%	\end{equation}
%}

Since each of the $M$ columns of $\hat{\mathbf{h}}^\mathrm{rad,corr}$ contains information on the range of the $H$ targets, performing row-wise \acp{DFT} on $\hat{\mathbf{h}}^\mathrm{rad,corr}$ converts the phases $\phi_{m,\eta}$ into Doppler shift information. The aforementioned processing produces the matrix $\mathbf{I}^\mathrm{rad}\in\mathbb{C}^{N\times M}$, which allows to clearly extract information on the range and relative radial velocity of all $H$ targets if they are resolved in at least one of the radar image dimensions. Based on the analyses in \cite{nuss2018,giroto2021_tmtt,giroto2020_OCDM_RadCom,giroto2021_OCDM_RadCom}, the aforementioned radar image experiences a processing gain $G_\text{p}$ and is associated with range resolution $\Delta R$ and maximum unambiguous range \mbox{$R_\text{max,ua}$}, as well as relative radial velocity resolution $\Delta v$ and maximum unambiguous relative radial velocity $v_\text{max,ua}$ values as shown in Table~\ref{tab:RadarParameters}. 
Since all transmit \ac{OCDM} symbols within the frame are equal and therefore no \ac{CP} is required, no further restriction on the maximum range value such as in \cite{giroto2021_tmtt,giroto2020_OCDM_RadCom}, and \cite{giroto2021_OCDM_RadCom} is considered. Furthermore, since a restriction on the maximum tolerable relative radial velocity for the considered \ac{OCDM}-based system will be accurately investigated in Section~\ref{sec:results}, it is not listed in Table~\ref{tab:RadarParameters}.

\begin{table}[!t]
	\renewcommand{\arraystretch}{1.5}
	\arrayrulecolor[HTML]{708090}
	\setlength{\arrayrulewidth}{.1mm}
	\setlength{\tabcolsep}{4pt}
	
	\centering
	\caption{Performance parameters in an OCDM-based radar system with discrete-Fresnel domain channel estimation. If all OCDM symbols within the frame are equal, no CP is needed and therefore $N_\mathrm{CP}$ can be set to zero.}
	\label{tab:RadarParameters}
	%\tiny
	\begin{tabular}{|cc|}
		\hhline{|==|}
		\multicolumn{2}{|c|}{\textbf{Radar performance parameters}} \\ \hhline{|==|}
		\multicolumn{1}{|c|}{\textbf{Processing gain}}      & $G_\mathrm{p} = NM$ \\ \hline
		\multicolumn{1}{|c|}{\textbf{Range resolution}}     & $\Delta R = c_0/(2B)$ \\ \hline
		\multicolumn{1}{|c|}{\textbf{Max. unamb. range}}    & $R_\mathrm{max,ua} = N~c_0/(2B)$ \\ \hline
		\multicolumn{1}{|c|}{\textbf{Velocity resolution}}  & $\Delta v = B~c_0/\left[2f_\text{c}\left(N+N_\mathrm{CP}\right)M\right]$ \\ \hline
		\multicolumn{1}{|c|}{\textbf{Max. unamb. velocity}} & $v_\mathrm{max,ua} = B~c_0/\left[4f_\text{c}\left(N+N_\mathrm{CP}\right)\right]$ \\ \hhline{|==|}
	\end{tabular}
\end{table}

\subsection{Extension to MU or MIMO Radar Operation}\label{subsec:multiplexing}

\begin{figure}[!t]
	\centering
	
	\psfrag{A}[c][c]{\tiny $N/P$}
	\psfrag{1}[c][c]{\scriptsize $0$}
	\psfrag{2}[c][c]{\scriptsize $1$}
	\psfrag{3}[c][c]{\scriptsize $P-1$}
	
	\psfrag{B}[c][c]{\tiny $0$}
	\psfrag{C}[c][c]{\tiny $N/P$}
	\psfrag{D}[c][c]{\tiny $N-N/P$}
	\psfrag{E}[c][c]{\tiny $N-1$}
	
	\psfrag{H}[c][c]{\tiny $k$}	
	
	\includegraphics[width=8.5cm]{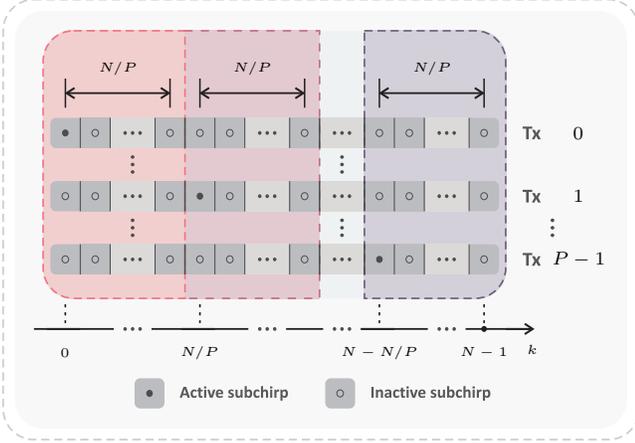}\caption{Subchirp allocation in the discrete-Fresnel domain among distinct transmit channels for MU or MIMO operation based on FrDM.}\label{fig:extensionMIMO}
	%\vspace{-0.25cm}
\end{figure}

If \mbox{$P\in\mathbb{N}_+$} synchronized \ac{OCDM} radar transmitters labeled as \mbox{$p\in\{0,1,\dots,P-1\}$} are to operate simultaneously in a \ac{MU} or \ac{MIMO} scenario, then the elements of the matrix \mbox{$\mathbf{\dot{X}}^p\in\mathbb{C}^{N\times M}$} that represents the discrete-Fresnel domain transmit frame of the $p\mathrm{th}$ transmitter are given by
\begin{equation}\label{eq:FresnelPilotMIMO}
	\dot{X}^p_{k,m} = \delta[\left<k - pN/P\right>_N].
\end{equation}
This equation indicates that the \ac{FrDM} scheme \cite{giroto2020_OCDM_RadCom}, which was originally proposed in \cite{giroto2021} for distributed reflectometric sensing of power lines, is adopted to enable orthogonal transmission of signals associated with each of the $P$ transmitters as depicted in Fig.~\ref{fig:extensionMIMO}.

Assuming that there are \mbox{$Q\in\mathbb{N}_+$} synchronized \ac{OCDM} radar receivers labeled as \mbox{$q\in\{0,1,\dots,Q-1\}$} and following the formulation in Section~\ref{subsec:radarChannelEst}, the elements of the matrix \mbox{$\mathbf{\dot{Y}}^q\in\mathbb{C}^{N\times M}$} representing the receive frame of the $q\mathrm{th}$ receiver are given by
\begin{equation}
	\dot{Y}^{q}_{k,m} = \sum\limits_{p=0}^{P-1} \hat{h}^{\mathrm{rad,corr},p,q}_{\left<k- pN/P\right>_N,m},
\end{equation}
where \mbox{$\hat{h}^{\mathrm{rad,corr},p,q}_{k,m}\in\mathbb{C}$} is the element at the $k\mathrm{th}$ row and $m\mathrm{th}$ column of the matrix \mbox{$\hat{\mathbf{h}}^{\mathrm{rad,corr},p,q}\in\mathbb{C}^{N\times M}$} that represents consectutive \ac{CIR} estimates associated with the $p\mathrm{th}$ transmitter and $q\mathrm{th}$ receiver. If the \acp{CIR} in \mbox{$\hat{\mathbf{h}}^{\mathrm{rad,corr},p,q}$} are assumed to have length equal to or smaller than $N/P$, i.e., $\hat{\mathbf{h}}^{\mathrm{rad,corr},p,q}$ can be redefined as \mbox{$\hat{\mathbf{h}}^{\mathrm{rad,corr},p,q}\in\mathbb{C}^{N/P\times 1}$}, than its elements \mbox{$\hat{h}^{\mathrm{rad,corr},p,q}_{k',m}\in\mathbb{C}$}, \mbox{$k'\in\{0,1,\dots,N/P-1\}$}, can be extracted from \mbox{$\dot{\mathbf{Y}}^q$} as
\begin{equation}
	\hat{h}^{\mathrm{rad,corr},p,q}_{k',m} = \dot{Y}^{q}_{k'+pN/P,m}.
\end{equation}

With the described \ac{FrDM} approach for enabling \ac{MU} or \ac{MIMO} radar operation, most performance parameters from Table~\ref{tab:RadarParameters} are kept as all \ac{OCDM} symbols in the transmit frame are equal and \acp{CP} are not needed. The only exception is the maximum unambiguous range, which becomes $R^{\mathrm{MU/MIMO},P}_\mathrm{max,ua} = (N/P)~c_0/(2B)$. It is worth highlighting that, although the ultimatelly obtained \ac{CIR} estimates have the reduced length of $N/P$, the processing gain remains proportional to $N$ since it is already experienced after the \ac{DFnT} at the receiver, which performs a pulse-compression like operation on the $N$ discrete-time domain samples and takes place before the $N/P$ samples associated with each transmitter are selected at each receiver.

%-----------------------------------------------------------%

\begin{figure*}[!b]
		%\vspace*{4pt}
		\hrulefill
		%\vspace*{1pt}
		\setcounter{equation}{40}
		\begin{equation}\label{eq:Y_RadCom_radar}
		\dot{Y}^{\mathrm{rad},q}_{k,m} = \hat{h}^\mathrm{rad,corr}_{k,m} + \sum\limits_{k''=0}^{(N-2N_\mathrm{CP}+1)-1}C_{k'',m}\sum\limits_{\eta=0}^{H-1}\e^{\im \phi_{m,\eta}}\delta[\left<k-(n_{\Delta,\eta}+N_\mathrm{CP}+k_{\Delta,\eta})\right>_N]~\e^{\im\frac{\pi}{N}\left[2nk_{\Delta,\eta}-k_{\Delta,\eta}^2+N(n_{\Delta,\eta}k_{\Delta,\eta})\right]}
		\end{equation}
\end{figure*}

\begin{figure}[!t]
	\centering
	
	\psfrag{A}[c][c]{\tiny $N_\mathrm{CP}$}
	\psfrag{B}[c][c]{\tiny $N-2N_\mathrm{CP}+1$}
	\psfrag{C}[c][c]{\tiny $N_\mathrm{CP}-1$}
	
	\psfrag{D}[c][c]{\tiny $0$}
	\psfrag{E}[c][c]{\tiny $N_\mathrm{CP}$}
	\psfrag{F}[c][c]{\tiny $N-N_\mathrm{CP}+1$}
	\psfrag{G}[c][c]{\tiny $N-1$} 
	
	\psfrag{H}[c][c]{\tiny $k$}	
	
	\includegraphics[width=8.5cm]{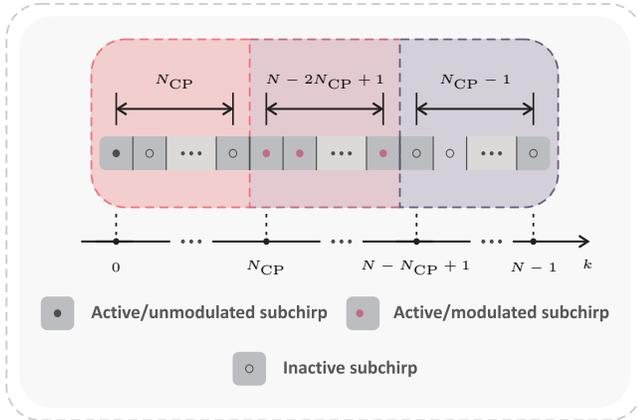}\caption{Subchirp allocation in the discrete-Fresnel domain for enabling joint RadCom operation. The first active, unmodulated subchirp and the next $N_\mathrm{CP}-1$ inactive subchirps are used for radar sensing, while the \mbox{$N-2N_\mathrm{CP}+1$} active, modulated subchirps are used for comunication. The last \mbox{$N_\mathrm{CP}-1$} inactive subchirps are used as a guard interval.}\label{fig:extensionRadCom}
	%\vspace{-0.25cm}
\end{figure}

%-----------------------------------------------------------%

\subsection{Extension to RadCom Operation}\label{subsec:radCom}

To enable joint \ac{RadCom} operation of an \ac{OCDM}-based radar system with discrete-Fresnel domain estimation, the \ac{FrDM} principle explained in Section~\ref{subsec:multiplexing} is applied. In this context, a sector-modulated \ac{OCDM} symbol structure in the discrete-Fresnel domain based on the autocorrelation pattern of \ac{ZCZ} sequences used in \ac{PMCW} radars is adopted. Unlike in the radar-only case from Section~\ref{subsec:radarChannelEst}, each of the $M$ \ac{OCDM} symbols in the transmit frame will be unique as they carry different modulated data. Consequently, a \ac{CP} of length \mbox{$N_\mathrm{CP}\in\mathbb{N}_+|\{2N_\mathrm{CP}-1<N\}$} must be appended to each \ac{OCDM} symbol in the discrete-time domain to avoid \ac{ISI}. The resulting maximum tolerable range in the considered \ac{OCDM}-based \ac{RadCom} system is $R_\mathrm{max,CP} = N_\mathrm{CP}~c_0/(2B)$, which is assumed to be lower than the maximum unambiguous range $R_\mathrm{max,ua}$ listed in Table~\ref{tab:RadarParameters}.

For a given $N_\mathrm{CP}$, the \ac{FrDM}-based design of the \ac{OCDM} symbol depicted in Fig.~\ref{fig:extensionRadCom} is adopted and described as follows. First, the active, unmodulated subchirp of index $k=0$ from \eqref{eq:FresnelPilot} is kept and allocated and energy $\mathcal{E}^\mathrm{Rad}\in\mathbb{R}_+$, being followed by \mbox{$N_\mathrm{CP}-1$} inactive subchirps. At the receiver side, these first $N_\mathrm{CP}$ elements of the discrete-Fresnel domain \ac{OCDM} symbol will ultimately contain the radar \ac{CIR} estimate. Next, subchirps $k=N_\mathrm{CP}$ to $k=N-N_\mathrm{CP}+1$, which comprise a total of $N-2N_\mathrm{CP}+1$ subchirps, are modulated with symbols belonging to a digital modulation constellation that are contained in the matrix $\mathbf{C}\in\mathbb{C}^{\left(N-2N_\mathrm{CP}+1\right)\times M}$. Finally, the last $N_\mathrm{CP}-1$ are kept inactive, which, taking into account energy factors unlike in the previous sections, yields a discrete-Fresnel domain transmit frame whose elements are given by
\setcounter{equation}{39}
\begin{align}\label{eq:X_frame_RadCom}
	\dot{X}_{k,m} = &\sqrt{\mathcal{E}^\mathrm{Rad}}\delta[k]\nonumber\\
	&+ \sum\limits_{k''=0}^{(N-2N_\mathrm{CP}+1)-1} C_{k'',m}\delta[\left<k-(N_\mathrm{CP}+k'')\right>_N].
\end{align}
In this equation, \mbox{$k''\in\{0,1,\dots,(N-2N_\mathrm{CP}+1)-1\}$} and $C_{k'',m}$ denotes an element of $\mathbf{C}\in\mathbb{C}^{\left(N-2N_\mathrm{CP}+1\right)\times M}$, which belongs to a digital modulation constellation with mean constellation energy that can be expressed as $\mathcal{E}^\mathrm{Com} = \mathbb{E}\{|C_{k'',m}|^2\}$ if $C_{k'',m}$ is regarded as a random variable. The total energy allocated to the transmit \ac{OCDM} symbol for joint \ac{RadCom} operation is consequently \mbox{$\mathcal{E}^\mathrm{RadCom}=\mathcal{E}^\mathrm{Rad}+(N-2N_\mathrm{CP}+1)\mathcal{E}^\mathrm{Com}$}. The last \mbox{$N_\mathrm{CP}-1$} inactive subchirps of each \ac{OCDM} symbol are used as a guard interval to avoid interference of the \mbox{$N-2N_\mathrm{CP}+1$} communication subchirps onto the $N_\mathrm{CP}$ radar subchirps, as the \ac{OCDM} symbols experience a circular convolution with the radar \acp{CIR}. Due to the \ac{OCDM} symbol structure from \eqref{eq:X_frame_RadCom}, the proposed \ac{OCDM}-based \ac{RadCom} system is named sector-modulated \ac{OCDM}-based \ac{RadCom} system.

After transmitted, the \ac{OCDM} \ac{RadCom} signal will not only reflect off radar targets and be received by the same \ac{OCDM} \ac{RadCom} system that originally transmitted it, but also propagate towards another \ac{OCDM} communication or \ac{RadCom} device. The radar case is first explained as follows. Assuming \mbox{$n_{\Delta,\eta}\in\mathbb{Z}$} and \mbox{$k_{\Delta,\eta}\in\mathbb{Z}$} for the sake of conciseness and using the result from \eqref{eq:timeFreqConv_rad3}, the elements of the matrix \mbox{$\mathbf{\dot{Y}}^\mathrm{rad}\in\mathbb{C}^{N\times M}$} representing the radar receive frame after \ac{CP} removal and transformation into the discrete-Fresnel domain are expressed as in \eqref{eq:Y_RadCom_radar}. Similarly, the result for \mbox{$n_{\Delta,\eta}\in\mathbb{R}$} and \mbox{$k_{\Delta,\eta}\in\mathbb{R}$} can be achieved based on the result from \eqref{eq:timeFreqConv_rad4}. Since the length of the \acp{CIR} in \mbox{$\hat{\mathbf{h}}^\mathrm{rad}$} is equal to or smaller than $N_\mathrm{CP}$, the aforementioned \ac{CIR} matrix can be redefined as \mbox{$\hat{\mathbf{h}}^\mathrm{rad}\in\mathbb{C}^{N_\mathrm{CP}\times M}$}. Based on both \eqref{eq:Y_RadCom_radar} and the sufficient guard interval composed by null subchirps at the end of the discrete-Fresnel domain \ac{OCDM} symbols in \eqref{eq:X_frame_RadCom}, the elements \mbox{$\hat{h}^\mathrm{rad,corr}_{k''',m}\in\mathbb{C}$}, \mbox{$k'''\in\{0,1,\dots,N_\mathrm{CP}-1\}$}, of the aforementioned \ac{CIR} matrix can be extracted from \mbox{$\dot{\mathbf{Y}}^{\mathrm{rad}}\in\mathbb{C}^{N\times M}$} as
\setcounter{equation}{41}
\begin{equation}\label{eq:}
	\hat{h}^{\mathrm{rad,corr}}_{k''',m} \approx \dot{Y}^\mathrm{rad}_{k''',m}.
\end{equation}
Apart from the aforementioned maximum range, all other radar performance parameters from Table~\ref{tab:RadarParameters} also hold for the considered sector-modulated \ac{OCDM}-based \ac{RadCom} system with discrete-Fresnel domain channel estimation.

%As for the captured signal by the receiver of another \ac{OCDM} communication or \ac{RadCom} device, it is henceforth assumed that both timing and frequency synchronizations are ensured via techniques such as Schmidl \& Cox's algorithm \cite{schmidl1997,ouyang2017_diss}. After \ac{CP} removal and transformation into the discrete-Fresnel domain, \ac{CIR} estimates can be extracted from the first $N_\mathrm{CP}$ elements of each column of the matrix \mbox{$\mathbf{\dot{Y}}^\mathrm{com}\in\mathbb{C}^{N\times M}$} that represents the discrete-Fresnel domain communication receive frame. After channel equalization, the modulation symbols contained in subchirps $k=N_\mathrm{CP}$ to $k=N-N_\mathrm{CP}+1$ can finally be extracted and demodulated following a typical \ac{OCDM} proecessing chain.
As for the captured signal by the receiver of another \ac{OCDM} communication or \ac{RadCom} device, it is henceforth assumed that both timing and frequency synchronizations are ensured via techniques such as Schmidl \& Cox's algorithm \cite{schmidl1997,ouyang2017_diss}. After \ac{CP} removal and transformation into the discrete-Fresnel domain, \ac{CIR} estimates can be obtained from the same unmodulated pilot subchirps that used for radar sensing as previously mentioned. After channel equalization, the modulation symbols contained in subchirps $k=N_\mathrm{CP}$ to $k=N-N_\mathrm{CP}+1$ can finally be extracted and demodulated following a typical \ac{OCDM} processing chain.

%***********************************************************%
%-----------------------------------------------------------%
%***********************************************************%

\section{Numerical and Measurement Results}\label{sec:results}

In this section, a performance analysis of discrete-Fresnel domain channel estimation for \ac{OCDM}-based radar systems is performed. Aiming at mid-range \ac{HAD} applications \cite{giroto2021_tmtt}, a carrier frequency $f_\mathrm{c}=\SI{79}{\giga\hertz}$ and a frequency bandwidth $B=\SI{1}{\giga\hertz}$ are adopted. Additionally, $N=2048$ subchirps, no \ac{CP}, i.e., $N_\mathrm{CP}=0$, and $M=5120$ \ac{OCDM} symbols are considered unless explicitly stated otherwise, which results in an \ac{OCDM} symbol time duration of $\SI{2.05}{\micro\second}$, an \ac{OCDM} frame time duration of $\SI{10.49}{\milli\second}$, and in the radar performance parameters listed in Table~\ref{tab:RadarParameters_results}.

%Based on the aforementioned parameters, Fig.~\ref{fig:NRMSE} shows the simulated \ac{NRMSE} of the \ac{CIR} magnitude assuming a single target for a \ac{SISO} \ac{OCDM}-based radar performing discrete-Fresnel domain channel estimation. For the sake of simplicity, the subindex $\eta$ is henceforth omitted for $n_{\Delta,\eta}$ and $k_{\Delta,\eta}$. In this figure, the whole unambiguous interval for range and relative radial velocity were considered, which results in \mbox{$n_\Delta\in[0,2048]$} and \mbox{$k_\Delta\in[-0.5,0.5]$}. The values for $k_\Delta$ were defined based on the maximum unambiguous velocity expression from Table~\ref{tab:RadarParameters}, the relationship \mbox{$f_\mathrm{D}=2v/\lambda=2vf_\mathrm{c}/c_0$} between Doppler shifts and relative radial velocities, and the frequency resolution \mbox{$\Delta f=B/N$} of the \ac{OCDM}-based radar system. Consequently, $k_\Delta$ can be interpreted as a normalized Doppler shift expressed as \mbox{$k_\Delta=f_\mathrm{D}/\Delta f$}. It is observed in Fig.~\ref{fig:NRMSE} that the \ac{NRMSE} significantly increases approximately after $\left|k_\Delta\right|>0.1$, which is also expected, e.g., in \ac{OFDM}-based radar and \ac{RadCom} systems where the maximum tolerable velocity is associated with this $k_\Delta$ upper bound \cite{nuss2018,giroto2021_tmtt}.

\begin{table}[!t]
	\renewcommand{\arraystretch}{1.5}
	\arrayrulecolor[HTML]{708090}
	\setlength{\arrayrulewidth}{.1mm}
	\setlength{\tabcolsep}{4pt}
	
	\centering
	\caption{Resulting radar performance parameters in the considered OCDM-based radar system.}
	\label{tab:RadarParameters_results}
	%\tiny
	\begin{tabular}{|cc|}
		\hhline{|==|}
		\multicolumn{2}{|c|}{\textbf{Radar performance parameters}} \\ \hhline{|==|}
		\multicolumn{1}{|c|}{\textbf{Processing gain}}      & $G_\mathrm{p} = \SI{70.21}{\mathrm{dB}}$ \\ \hline
		\multicolumn{1}{|c|}{\textbf{Range resolution}}     & $\Delta R = \SI{0.15}{\meter}$ \\ \hline
		\multicolumn{1}{|c|}{\textbf{Max. unamb. range}}    & $R_\mathrm{max,ua} = \SI{307.20}{\meter}$ \\ \hline
		\multicolumn{1}{|c|}{\textbf{Velocity resolution}}  & $\Delta v = \SI{0.18}{\meter/\second}$ \\ \hline
		\multicolumn{1}{|c|}{\textbf{Max. unamb. velocity}} & $v_\mathrm{max,ua} = \SI{463.56}{\meter/\second}$ \\ \hhline{|==|}
	\end{tabular}
\end{table}
%\begin{figure}[!t]
%	\centering
%	
%	\psfrag{0}[c][c]{\small $0$}
%	\psfrag{255}[c][c]{\small $255$}
%	\psfrag{555}[c][c]{\small $511$}
%	\psfrag{767}[c][c]{\small $767$}
%	\psfrag{1023}[c][c]{\small $1023$}
%	\psfrag{1279}[c][c]{\small $1279$}
%	\psfrag{1535}[c][c]{\small $1535$}
%	\psfrag{1791}[c][c]{\small $1791$}
%	\psfrag{2047}[c][c]{\small $2047$}
%	
%	\psfrag{AAA}[c][c]{\small -$0.5$}
%	\psfrag{BBB}[c][c]{\small -$0.25$}
%	\psfrag{CCC}[c][c]{\small $0$}
%	\psfrag{DDD}[c][c]{\small $0.25$}
%	\psfrag{EEE}[c][c]{\small $0.5$}
%	
%	\psfrag{0.2}[c][c]{\small $0.2$}
%	\psfrag{0.4}[c][c]{\small $0.4$}
%	\psfrag{0.6}[c][c]{\small $0.6$}
%	\psfrag{0.8}[c][c]{\small $0.8$}
%	\psfrag{1}[c][c]{\small $1$}
%	
%	\psfrag{X}{$k_\Delta$}
%	\psfrag{Y}{$n_\Delta$}
%	\psfrag{xNRMSExx}{$\mathrm{NRMSE}$}
%	
%	\includegraphics[width=6cm]{./Figures/NRMSE_amplitude_interpCIR.eps}\caption{NRMSE of the magnitude of the estimated CIR assuming a single target for different $n_\Delta$ and $k_\Delta$ combinations.}\label{fig:NRMSE}
%	%\vspace{-0.25cm}
%\end{figure}

\begin{figure*}[!t]
	\centering
	
	\psfrag{55}{(a)}
	\psfrag{22}{(b)}
	\psfrag{33}{(c)}
	\psfrag{44}{(d)}
	
	\psfrag{0}[c][c]{\small $0$}
	\psfrag{255}[c][c]{\small $255$}
	\psfrag{555}[c][c]{\small $511$}
	\psfrag{767}[c][c]{\small $767$}
	\psfrag{1023}[c][c]{\small $1023$}
	\psfrag{1279}[c][c]{\small $1279$}
	\psfrag{1535}[c][c]{\small $1535$}
	\psfrag{1791}[c][c]{\small $1791$}
	\psfrag{2047}[c][c]{\small $2047$}
	
	\psfrag{AAA}[c][c]{\small -$0.5$}
	\psfrag{BBB}[c][c]{\small -$0.25$}
	\psfrag{CCC}[c][c]{\small $0$}
	\psfrag{DDD}[c][c]{\small $0.25$}
	\psfrag{EEE}[c][c]{\small $0.5$}
	
	\psfrag{-4}[c][c]{\small -$4$}
	\psfrag{-3}[c][c]{\small -$3$}
	\psfrag{-2}[c][c]{\small -$2$}
	\psfrag{-1}[c][c]{\small -$1$}
	\psfrag{0}[c][c]{\small $0$}
	
	\psfrag{-16}[c][c]{\small -$16$}
	\psfrag{-12}[c][c]{\small -$12$}
	\psfrag{-8}[c][c]{\small -$8$}
	\psfrag{-4}[c][c]{\small -$4$}
	\psfrag{0}[c][c]{\small $0$}
	
	\psfrag{-12}[c][c]{\small -$12$}
	\psfrag{-9}[c][c]{\small -$9$}
	\psfrag{-6}[c][c]{\small -$6$}
	\psfrag{-3}[c][c]{\small -$3$}
	\psfrag{0}[c][c]{\small $0$}
	\psfrag{3}[c][c]{\small $3$}
	
	\psfrag{X}{$k_\Delta$}
	\psfrag{Y}{$n_\Delta$}
	\psfrag{xPPLR (dB)xx}{$\mathit{PPLR}$ (dB)}
	\psfrag{xPSLR (dB)xx}{$\mathit{PSLR}$ (dB)}
	\psfrag{xISLR (dB)xx}{$\mathit{ISLR}$ (dB)}
	
	\psfrag{KKKK}{}
	\psfrag{KKKKKKKKKKK}{}
	
	\includegraphics[width=0.9\textwidth]{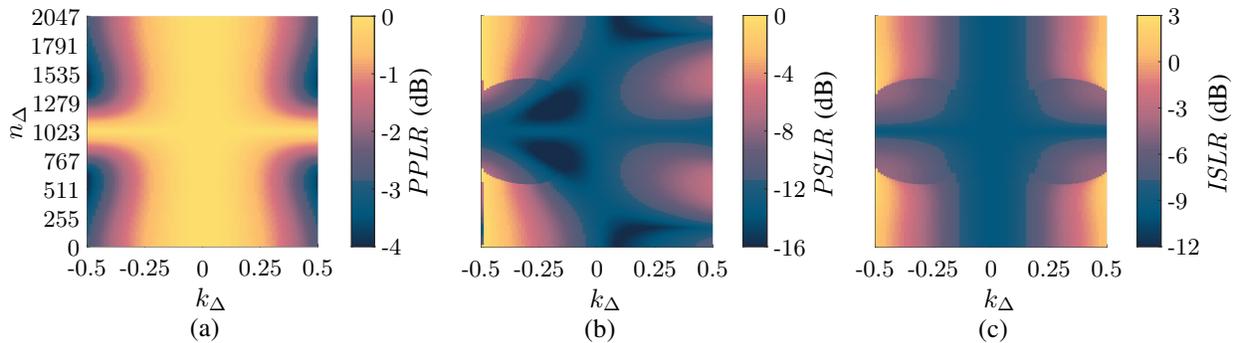}
	\caption{Simulated Doppler-shift tolerance: (a) PPLR, (b) PSLR, and (c) ISLR of a single target reflection with different \mbox{$n_{\Delta}=2RB/c_0$} and \mbox{$k_\Delta=f_\mathrm{D}/\Delta f$} pairs obtained in an \ac{OCDM}-based system with discrete-Fresnel domain radar processing.}\label{fig:DopplerRobustness}
	%\vspace{-0.25cm}
\end{figure*}

\begin{figure*}[!t]
	\centering
	
	\psfrag{55}{(a)}
	\psfrag{22}{(b)}
	\psfrag{33}{(c)}
	\psfrag{44}{(d)}
	
	\psfrag{AAAA}[c][c]{\small $29$}
	\psfrag{BBBB}[c][c]{\small $29.5$}
	\psfrag{CCCC}[c][c]{\small $30$}
	\psfrag{DDDD}[c][c]{\small $30.5$}
	\psfrag{EEEE}[c][c]{\small $31$}
	
	\psfrag{-2}[c][c]{\small -$2$}
	%\psfrag{-1}[c][c]{\small -$1$}
	\psfrag{-1}[c][c]{}
	\psfrag{0}[c][c]{\small $0$}
	%\psfrag{1}[c][c]{\small $1$}
	\psfrag{1}[c][c]{}
	\psfrag{2}[c][c]{\small $2$}
	
	\psfrag{90.71}[c][c]{\small $90.71$}
	%\psfrag{91.71}[c][c]{\small $91.71$}
	\psfrag{91.71}[c][c]{}
	\psfrag{92.71}[c][c]{\small $92.71$}
	%\psfrag{93.71}[c][c]{\small $93.71$}
	\psfrag{93.71}[c][c]{}
	\psfrag{94.71}[c][c]{\small $94.71$}
	
	\psfrag{229.78}[c][c]{\small $229.78$}
	%\psfrag{230.78}[c][c]{\small $230.78$}
	\psfrag{230.78}[c][c]{}
	\psfrag{231.78}[c][c]{\small $231.78$}
	%\psfrag{232.78}[c][c]{\small $232.78$}
	\psfrag{232.78}[c][c]{}
	\psfrag{233.78}[c][c]{\small $233.78$}
	
	\psfrag{461.56}[c][c]{\small $461.56$}
	%\psfrag{462.56}[c][c]{\small $462.56$}
	\psfrag{462.56}[c][c]{}
	\psfrag{463.56}[c][c]{\small $\pm463.56$}
	%\psfrag{464.56}[c][c]{\small $464.56$}
	\psfrag{464.56}[c][c]{}
	\psfrag{465.56}[c][c]{\small -$461.56$}
	
	\psfrag{0}[c][c]{\small $0$}
	\psfrag{-20}[c][c]{\small -$20$}
	\psfrag{-40}[c][c]{\small -$40$}
	\psfrag{-60}[c][c]{\small -$60$}
	
%	\psfrag{AAA}[c][c]{\small $28$}
%	\psfrag{BBB}[c][c]{\small $29$}
%	\psfrag{CCC}[c][c]{\small $30$}
%	\psfrag{DDD}[c][c]{\small $31$}
%	\psfrag{EEE}[c][c]{\small $32$}
	
	\psfrag{XXXXXXXXXXXXXXXXXXX}{\hspace{-0.075cm}Rel. radial velocity (m/s)}
	%\psfrag{Rel. radial velocity (m/s)}{\hspace{-0.075cm}Rel. radial velocity (m/s)}
	\psfrag{YYYYYYYYYYY}{\hspace{0.175cm}Range (m)}
	\psfrag{ZZZZZZZZZZZZZZZZZZZZZZZZZZZZZ}{\hspace{0.65cm}Norm. mag. (dB)}
	%\psfrag{xNorm. mag. (dB)xx}{\hspace{0.65cm}Norm. mag. (dB)}
	
	\psfrag{FFFF}{}
	\psfrag{FFFFFFFFFFF}{}
	
	\includegraphics[width=\textwidth]{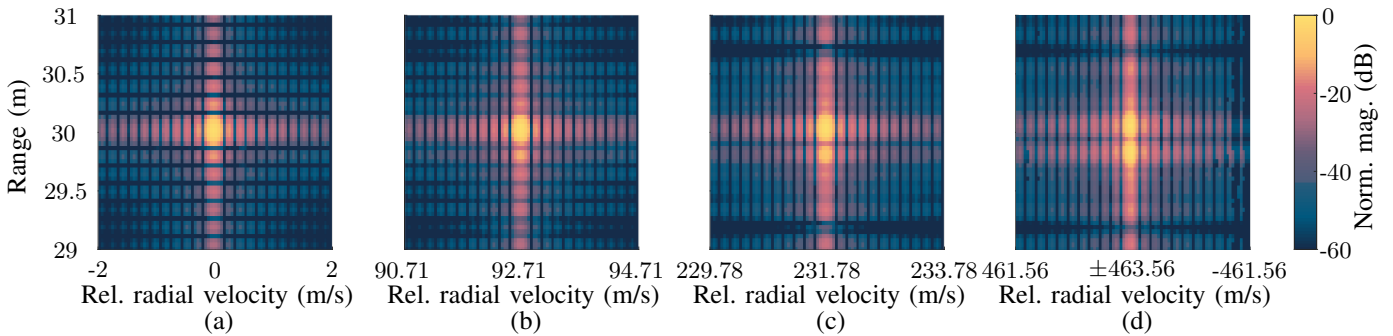}
	\caption{Obtained range-velocity radar images from measurements with OCDM-based radar systems with discrete-Fresnel domain channel estimation for a single target with relative radial velocities of (a)~$\SI{0}{\meter/\second}$ ($k_\Delta=0$), (b)~$\SI{92.71}{\meter/\second}$ ($k_\Delta=-0.1$), (c)~$\SI{231.78}{\meter/\second}$ ($k_\Delta=-0.25$), and (d)~$\SI{463.56}{\meter/\second}$ ($k_\Delta=-0.5$). In all radar images the actual target range is $\SI{30}{\meter}$ ($n_\Delta=200$).}\label{fig:RadarImages_4}
	%\vspace{-0.25cm}
\end{figure*}

Based on the aforementioned parameters, simulated \ac{PPLR}, \ac{PSLR}, and \ac{ISLR} \cite{lellouch2016,giroto2021_tmtt} results are shown in Fig.~\ref{fig:DopplerRobustness} to assess the distortion of the range mainlobe and sidelobes induced by Doppler shifts. Since those calculations assume a single point target, the subindex $\eta$ is henceforth omitted for $n_{\Delta,\eta}$ and $k_{\Delta,\eta}$ for the sake of simplicity. In this figure, the whole unambiguous interval for range and relative radial velocity were considered, which results in \mbox{$n_\Delta\in[0,2048]$} and \mbox{$k_\Delta\in[-0.5,0.5]$}. The values for $k_\Delta$ were defined based on the maximum unambiguous velocity expression from Table~\ref{tab:RadarParameters}, the relationship \mbox{$f_\mathrm{D}=2v/\lambda=2vf_\mathrm{c}/c_0$} between Doppler shifts and relative radial velocities, and the frequency resolution \mbox{$\Delta f=B/N$} of the \ac{OCDM}-based radar system. Consequently, $k_\Delta$ can be interpreted as a normalized Doppler shift expressed as \mbox{$k_\Delta=f_\mathrm{D}/\Delta f$}. Similarly, $n_\Delta$ can be interpreted as a normalized range by the range resolution and expressed as \mbox{$n_\Delta=R/\Delta R$}. Overall, degradations of \ac{PPLR}, \ac{PSLR}, and \ac{ISLR} are mostly only observed for $\left|k_\Delta\right|>0.1$, which is also expected, e.g., in \ac{OFDM}-based radar and \ac{RadCom} systems where the maximum tolerable velocity is associated with this $k_\Delta$ upper bound \cite{nuss2018,giroto2021_tmtt}. The \ac{PPLR} results in Fig.~\ref{fig:DopplerRobustness}(a) show a highest degradation of the processing gain $G_\mathrm{p}$ of up to about $\SI{4}{dB}$, e.g., for the approximate region covered by \mbox{$n_\Delta\in[0, 767]\cup[1279,2048]$} and \mbox{$k_\Delta=-0.5$}. Next, Fig.~\ref{fig:DopplerRobustness}(b) shows the \ac{PSLR} results, which indicate the ratio between the highest sidelobe and the mainlobe, while the \ac{ISLR} results, which indicate the ratio between the integrated sidelobe level and the integrated mainlobe level, are presented in Fig.~\ref{fig:DopplerRobustness}(c). Near the aforementioned region in the \ac{PPLR} case, both \ac{PSLR} and \ac{ISLR} experience significant degradation, which can be explained by the results from \eqref{eq:timeFreqConv_rad3} and \eqref{eq:timeFreqConv_rad4}. Besides the range-Doppler coupling in these equations, the range and Doppler dependent phase terms, which are \mbox{$\e^{\im\frac{\pi}{N}\left(2nk_{\Delta,\eta}-k_{\Delta,\eta}^2+Nk_{\Delta,\eta}\right)}$} in \eqref{eq:timeFreqConv_rad4}, introduce changes in the shape of the estimated \acp{CIR} w.r.t. the ideally expected ones and is illustrated in more detail as follows.

Fig.~\ref{fig:RadarImages_4} shows range-velocity radar images obtained from measurements with a Zynq UltraScale+ RFSoC ZCU111 from Xilinx, Inc. The aforementioned \ac{SoC} platform was used to emulate both the \ac{OCDM}-based radar system with the adopted parameters in this section and transmit power \mbox{$P_\text{Tx}=\SI{0}{dBm}$}, as well as the \ac{RTS} described in \cite{diewald2021_journal1} for a single radar target with \ac{RCS} \mbox{$\sigma_\mathrm{RCS}=\SI{30}{dBsm}$} at $\SI{30}{\meter}$ ($n_\Delta=200$) and velocities ranging from $\SI{00}{\meter/\second}$ ($k_\Delta=0$) to $\SI{463.56}{\meter/\second}$ ($k_\Delta=-0.5$). In these images, it is possible to see that, for increasing radial relative velocities and consequently Doppler shifts, the aforementioned range-Doppler coupling and the range and Doppler dependent phase terms may bias the target reflection from its actual range and duplicate the target's reflection in the range direction if $k_\Delta=0$, which explains the \ac{PPLR}, \ac{PSLR}, and \ac{ISLR} results from Fig.~\ref{fig:DopplerRobustness}. Conversely, the Doppler shift or relative radial velocity estimation is not affected by the aforementioned effects and only depends on the Doppler-induced phases $\e^{\im \phi_{m,\eta}}$ in \eqref{eq:timeFreqConv_rad3} and \eqref{eq:timeFreqConv_rad4}. As a consequence, similar performance for the velocity processing, e.g., to \ac{OFDM} or \ac{PMCW} radars is achieved.

\begin{figure*}[!t]
	\centering
	
	\psfrag{55}{(a)}
	\psfrag{22}{(b)}
	\psfrag{33}{(c)}
	\psfrag{44}{(d)}
	
	\psfrag{0}[c][c]{\small $0$}
	\psfrag{-10}[c][c]{\small -$10$}
	\psfrag{-20}[c][c]{\small -$20$}
	\psfrag{-30}[c][c]{\small -$30$}
	\psfrag{-40}[c][c]{\small -$40$}
	
	\psfrag{28.80}[c][c]{\small $28.80$}
	\psfrag{28.95}[c][c]{\small $28.95$}
	\psfrag{29.10}[c][c]{\small $29.10$}
	\psfrag{29.25}[c][c]{\small $29.25$}
	\psfrag{29.40}[c][c]{\small $29.40$}
	\psfrag{29.55}[c][c]{\small $29.55$}
	\psfrag{29.70}[c][c]{\small $29.70$}
	\psfrag{29.85}[c][c]{\small $29.85$}
	\psfrag{30.00}[c][c]{\small $30.00$}
	\psfrag{30.15}[c][c]{\small $30.15$}
	\psfrag{30.30}[c][c]{\small $30.30$}
	\psfrag{30.45}[c][c]{\small $30.45$}
	\psfrag{30.60}[c][c]{\small $30.60$}
	\psfrag{30.75}[c][c]{\small $30.75$}
	\psfrag{30.90}[c][c]{\small $30.90$}
	\psfrag{31.05}[c][c]{\small $31.05$}
	\psfrag{31.20}[c][c]{\small $31.20$}
	
	\psfrag{AAA}[c][c]{\small -$0.5$}
	\psfrag{BBB}[c][c]{\small -$0.25$}
	\psfrag{CCC}[c][c]{\small $0$}
	\psfrag{DDD}[c][c]{\small $0.25$}
	\psfrag{EEE}[c][c]{\small $0.5$}
	
	\psfrag{0.2}[c][c]{\small $0.2$}
	\psfrag{0.4}[c][c]{\small $0.4$}
	\psfrag{0.6}[c][c]{\small $0.6$}
	\psfrag{0.8}[c][c]{\small $0.8$}
	\psfrag{1}[c][c]{\small $1$}
	
	\psfrag{Range (m)}{Range (m)}
	\psfrag{xNorm. mag. (dB)xx}{Norm. mag. (dB)}
	
	\includegraphics[width=0.9\textwidth]{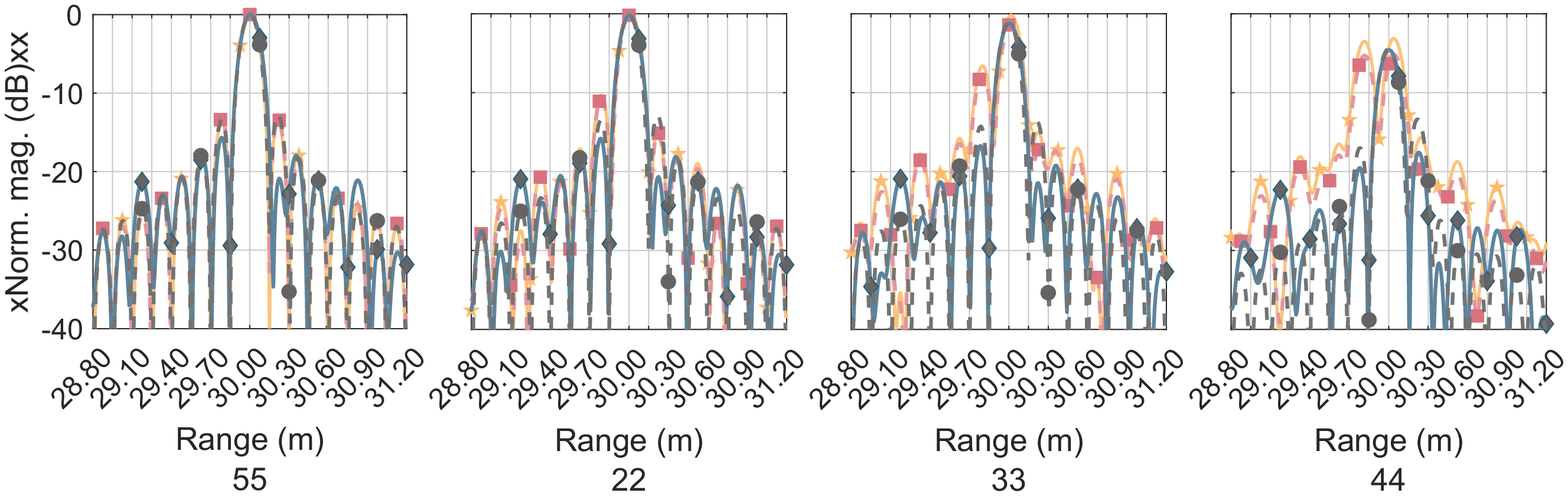}
	\caption{Range cuts of radar images from measurements with \ac{SISO} \ac{OCDM}-based radar system ({\color[rgb]{0.9882,0.7333,0.4275}$\bigstar$}), sector-modulated \ac{OCDM}-based \ac{RadCom} system ({\color[rgb]{0.8471,0.4510,0.4980}$\blacksquare$}), conventional \ac{OCDM}-based \ac{RadCom} system ({\color[rgb]{0.1490,0.3569,0.4824}$\blacklozenge$}), \ac{OFDM}-based \ac{RadCom} system ({\color[rgb]{0.3922,0.3922,0.3922}$\CIRCLE$}) for a single target with relative radial velocity of (a)~$\SI{0}{\meter/\second}$ ($k_\Delta=0$), (b)~$\SI{92.71}{\meter/\second}$ ($k_\Delta=-0.1$), (c)~$\SI{231.78}{\meter/\second}$ ($k_\Delta=-0.25$), and (d)~$\SI{463.56}{\meter/\second}$ ($k_\Delta=-0.5$). In all radar images the actual target range is $\SI{30}{\meter}$ ($n_\Delta=200$). Since the three latter modulation schemes yield processing gain around $\SI{1}{dB}$ lower than the \ac{OCDM}-based radar system, the magnitudes of the range profiles obtained with each modulation scheme are normalized w.r.t. their respective results at $\SI{0}{\meter/\second}$ ($k_\Delta=0$) for a fairer comparison.}\label{fig:Example_Range_profiles_SISO}
	%\vspace{-0.25cm}
\end{figure*}

For a more detailed visualization, Fig.~\ref{fig:Example_Range_profiles_SISO} shows range cuts at the estimated target velocities of the measured radar images from Fig.~\ref{fig:RadarImages_4} with a focus on the target's range. Results for three additional systems parameterized to yield the same frame duration are shown for each considered velocity in Fig.~\ref{fig:Example_Range_profiles_SISO} for the sake of comparison. The first are for the proposed sector-modulated \ac{OCDM}-based \ac{RadCom} system in Section~\ref{subsec:radCom} with  $N=2048$ subchirps, \ac{CP} length of $N_\mathrm{CP}=512$, and $M=4096$ \ac{OCDM} symbols. The second set of results is for the conventional \ac{OCDM}-based radar system from \cite{giroto2021_OCDM_RadCom} with  $N=2048$ subchirps, $N_\mathrm{CP}=512$ \ac{CP} length, and $M=4096$ \ac{OCDM} symbols. Finally, the third set of results is for an \ac{OFDM}-based \ac{RadCom} system with  $N=2048$ subcarriers, \ac{CP} length of $N_\mathrm{CP}=512$, and $M=4096$ \ac{OFDM} symbols. For all three aforementioned modulation schemes, the resulting maximum tolerable range defined by the \ac{CP} length is \mbox{$R_\mathrm{max,CP} = \SI{76.8}{\meter}$}, the achieved processing gain is $G_\mathrm{p} = \SI{69.24}{\mathrm{dB}}$, and \ac{QPSK} modulation with uniform power allocation is adopted for the modulated subchirps or subcarriers. As shown in Fig.~\ref{fig:Example_Range_profiles_SISO}, the proposed \ac{OCDM}-based radar system and its sector-modulated \ac{OCDM}-based \ac{RadCom} yield virtually equal radar range profiles. At zero Doppler shift, i.e., $k_\Delta=0$, they yield the same range profile result as their \ac{OFDM} counterpart and a more regular sidelobe trend than the conventional \ac{OCDM}-based \ac{RadCom} system from \cite{giroto2021_OCDM_RadCom}, whose range sidelobes are significantly influenced by the modulated data onto its subchirps due to its correlation-based range processing. As for $k_\Delta=-0.1$, which corresponds to the maximum tolerable Doppler shift for \ac{OFDM} \cite{sturm2011,nuss2018}, the proposed \ac{OCDM}-based radar and \ac{RadCom} systems present comparable performance to their \ac{OFDM} counterpart, while the conventional \ac{OCDM}-based \ac{RadCom} system from \cite{giroto2021_OCDM_RadCom} still yields the worse range sidelobe pattern. Relevant differences in the range profiles obtained with the proposed \ac{OCDM}-based systems w.r.t. \ac{OFDM} are only observed for $k_\Delta=-0.25$ and $k_\Delta=-0.5$, namely, peak splitting and noticeable range-Doppler coupling due to the effects described in \eqref{eq:timeFreqConv_rad3} and \eqref{eq:timeFreqConv_rad4}. However, it should be noted that these Doppler shift levels yield very high relative radial velocities and are therefore not to be expected in practical scenarios. Additionally, the combination of these Doppler shifts with the range of $\SI{30}{\meter}$ ($n_\Delta=200$) leads the proposed \ac{OCDM}-based radar and sector-modulated \ac{OCDM}-based \ac{RadCom} systems into their most critical region in terms of \ac{PPLR}, \ac{PSLR}, and \ac{ISLR} as shown in Fig.~\ref{fig:DopplerRobustness}. The results for these range and Doppler shift pairs are therefore lower bounds on the radar sensing performance of the proposed \ac{OCDM}-based schemes and better results are expected for less critical combinations of range and Doppler shift values.%However, this Doppler shift level yield very high relative radial velocities and are therefore not to be expected in practical scenarios.

\begin{figure*}[!t]
	\centering
	
	\psfrag{55}{(a)}
	\psfrag{22}{(b)}
	\psfrag{33}{(c)}
	\psfrag{44}{(d)}
	
	\psfrag{0}[c][c]{\small $0$}
	\psfrag{-10}[c][c]{\small -$10$}
	\psfrag{-20}[c][c]{\small -$20$}
	\psfrag{-30}[c][c]{\small -$30$}
	\psfrag{-40}[c][c]{\small -$40$}
	
	\psfrag{28.80}[c][c]{\small $28.80$}
	\psfrag{28.95}[c][c]{\small $28.95$}
	\psfrag{29.10}[c][c]{\small $29.10$}
	\psfrag{29.25}[c][c]{\small $29.25$}
	\psfrag{29.40}[c][c]{\small $29.40$}
	\psfrag{29.55}[c][c]{\small $29.55$}
	\psfrag{29.70}[c][c]{\small $29.70$}
	\psfrag{29.85}[c][c]{\small $29.85$}
	\psfrag{30.00}[c][c]{\small $30.00$}
	\psfrag{30.15}[c][c]{\small $30.15$}
	\psfrag{30.30}[c][c]{\small $30.30$}
	\psfrag{30.45}[c][c]{\small $30.45$}
	\psfrag{30.60}[c][c]{\small $30.60$}
	\psfrag{30.75}[c][c]{\small $30.75$}
	\psfrag{30.90}[c][c]{\small $30.90$}
	\psfrag{31.05}[c][c]{\small $31.05$}
	\psfrag{31.20}[c][c]{\small $31.20$}
	
	\psfrag{AAA}[c][c]{\small -$0.5$}
	\psfrag{BBB}[c][c]{\small -$0.25$}
	\psfrag{CCC}[c][c]{\small $0$}
	\psfrag{DDD}[c][c]{\small $0.25$}
	\psfrag{EEE}[c][c]{\small $0.5$}
	
	\psfrag{0.2}[c][c]{\small $0.2$}
	\psfrag{0.4}[c][c]{\small $0.4$}
	\psfrag{0.6}[c][c]{\small $0.6$}
	\psfrag{0.8}[c][c]{\small $0.8$}
	\psfrag{1}[c][c]{\small $1$}
	
	\psfrag{Range (m)}{Range (m)}
	\psfrag{xNorm. mag. (dB)xx}{Norm. mag. (dB)}
	
	\includegraphics[width=0.9\textwidth]{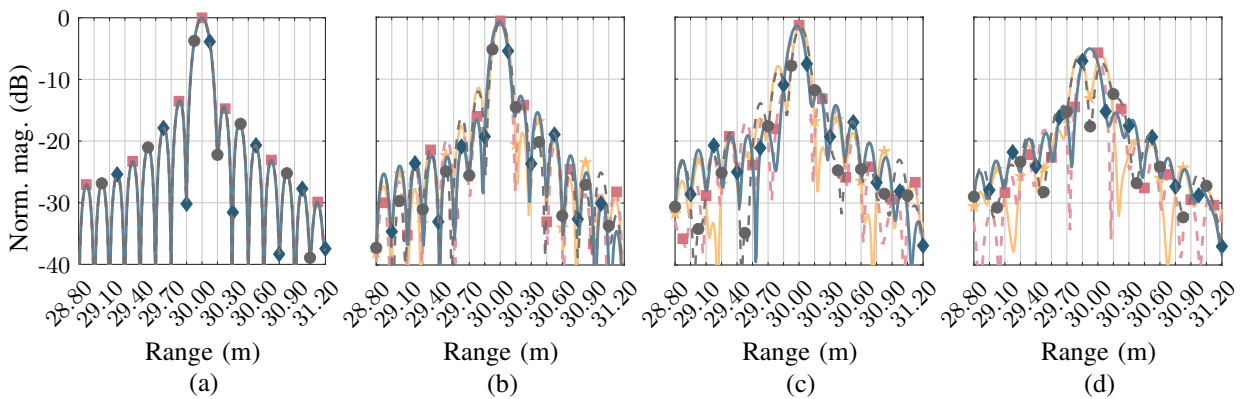}
	\caption{Range cuts of radar images from measurements with transmit channels $p=0$ ({\color[rgb]{0.9882,0.7333,0.4275}$\bigstar$}), $p=1$ ({\color[rgb]{0.8471,0.4510,0.4980}$\blacksquare$}), $p=2$ ({\color[rgb]{0.1490,0.3569,0.4824}$\blacklozenge$}), and $p=3$ ({\color[rgb]{0.3922,0.3922,0.3922}$\CIRCLE$}) in the considered \ac{MIMO} \ac{OCDM}-based radar system for a single target with relative radial velocity of (a)~$\SI{0}{\meter/\second}$ ($k_\Delta=0$), (b)~$\SI{92.71}{\meter/\second}$ ($k_\Delta=-0.1$), (c)~$\SI{231.78}{\meter/\second}$ ($k_\Delta=-0.25$), and (d)~$\SI{463.56}{\meter/\second}$ ($k_\Delta=-0.5$). In all radar images the actual target range is $\SI{30}{\meter}$ ($n_\Delta=200$).}\label{fig:Example_Range_profiles_MIMO}
	%({\color[rgb]{0.8471,0.4510,0.4980}$\blacksquare$}), OCDM-based RadCom system ({\color[rgb]{0.1490,0.3569,0.4824}$\blacklozenge$}) and an OFDM-based radar system ({\color[rgb]{0.3922,0.3922,0.3922}$\CIRCLE$})
	%\vspace{-0.25cm}
\end{figure*}

To address the \ac{FrDM} multiplexing strategy for \ac{MU} or \ac{MIMO} operation described in Section~\ref{subsec:multiplexing}, similar measurement results to the ones in Fig.~\ref{fig:Example_Range_profiles_SISO} are presented in Fig.~\ref{fig:Example_Range_profiles_MIMO} for signals transmitted by $P=4$ collocated transmitters and evaluated at a single receiver. For the results in this figure, the same \ac{OCDM} signal parameters adopted for the \ac{SISO} \ac{OCDM}-based radar system as mentioned at the beginning of this section are adopted, leading to the same radar performance parameters listed in Table~\ref{tab:RadarParameters_results}. The only exception is the maximum unambiguous range, which is reduced to \mbox{$R^{\mathrm{MU/MIMO},P}_\mathrm{max,ua} = \SI{76.8}{\meter}$} due to the multiplexing among $P=4$ transmitters. Compared to the results for the \ac{SISO} \ac{OCDM}-based radar system in Fig.~\ref{fig:Example_Range_profiles_SISO}, which are virtually the same as for the transmitter $p=0$ in the \ac{MIMO} case, the results from Fig.~\ref{fig:Example_Range_profiles_MIMO} only differ significantly for $k_\Delta=-0.25$ and $k_\Delta=-0.5$. This is due to the combined effect of the range-Doppler coupling and Doppler dependent phase terms in \eqref{eq:timeFreqConv_rad4}, which directly influence the peak splitting and range-Doppler coupling effects. As already mentioned previously, however, such high Doppler shift levels are not to be expected in practice.

\begin{figure}[!t]
	\centering
	
	\psfrag{Baseband frequency (GHz)}[c][c]{Baseband frequency (GHz)}
	\psfrag{Norm. CFR mag. (dB)}[c][c]{CFR magnitude (dB)}
	
	\psfrag{-0.5}[c][c]{\scriptsize -$0.5$}
	\psfrag{-0.375}[c][c]{\scriptsize -$0.375$}
	\psfrag{-0.25}[c][c]{\scriptsize -$0.25$}
	\psfrag{-0.125}[c][c]{\scriptsize -$0.125$}
	\psfrag{0}[c][c]{\scriptsize $0$}
	\psfrag{0.125}[c][c]{\scriptsize $0.125$}
	\psfrag{0.25}[c][c]{\scriptsize $0.25$}
	\psfrag{0.375}[c][c]{\scriptsize $0.375$}
	\psfrag{0.5}[c][c]{\scriptsize $0.5$}
	
	\psfrag{-30}[c][c]{\scriptsize -$30$}
	\psfrag{-27.5}[c][c]{\scriptsize -$27.5$}
	\psfrag{-25}[c][c]{\scriptsize -$25$}
	\psfrag{-22.5}[c][c]{\scriptsize -$22.5$}
	\psfrag{-20}[c][c]{\scriptsize -$20$}
	\psfrag{-17.5}[c][c]{\scriptsize -$17.5$}
	\psfrag{-15}[c][c]{\scriptsize -$15$}
	
	\includegraphics[width=8cm]{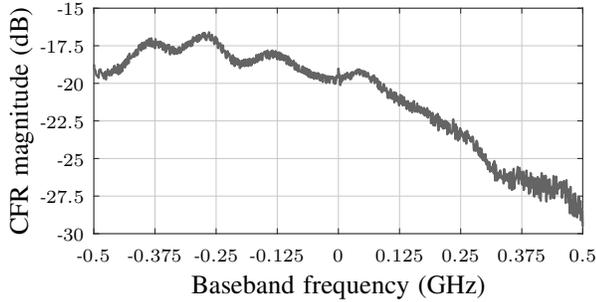}
	
	\caption{Magnitude response of the experienced communication \ac{CFR}.}\label{fig:H_CFR}
	
\end{figure}

To assess the communication performance of the proposed \ac{OCDM}-based \ac{RadCom} in Section~\ref{subsec:radCom}, measurements over a communication channel emulated by re-purposing the previously described \ac{RTS} were performed with the same \ac{SoC} platform from Xilinx, Inc, ensuring time and frequency synchronization between the communication transmitter and receiver with the \ac{SC} algorithm as described in \cite{filomeno2022}. For the communication analysis, the proposed sector-modulated \ac{OCDM}-based \ac{RadCom} system was compared with the \ac{OFDM}-based and conventional \ac{OCDM}-based \ac{RadCom} systems. Additionally, the same signal parameters adopted for the radar performance analysis were kept for all three modulation schemes, resulting in a data rate of $\SI{0.80}{\giga {bit}/\second}$ for the sector-modulated \ac{OCDM}-based \ac{RadCom} system and a data rate of $\SI{1.40}{\giga {bit}/\second}$ for its \ac{OFDM} and conventional \ac{OCDM} counterparts, which is influenced by inserted pilot subcarriers in the \ac{OFDM} case and use of \ac{FSP}-inserted pilots in the conventional \ac{OCDM} case \cite{omar2022}. The magnitude response of the experienced communication \ac{CFR}, whose frequency-selectivity is due to the sinc-shaping resulting from the use of an \ac{IF}-sampling architecture and uncalibrated effects of cables, baluns and filters, is shown in Fig.~\ref{fig:H_CFR}. To enable channel equalization and subsequent extraction of the receive \ac{QPSK} symbols from the receive signal, the unmodulated subchirp in the sector-modulated \ac{OCDM}-based \ac{RadCom} system was used. As for \ac{OFDM}- and the conventional \ac{OCDM}-based \ac{RadCom} system from \cite{giroto2020_OCDM_RadCom}, channel estimation was performed using the comb-pilot-based techniques, namely the one described in \cite{sit2018} for \ac{OFDM} and its \ac{FSP}-based version for the conventional \ac{OCDM}-based \ac{RadCom} system \cite{omar2022}. For all three modulation schemes, multiple channel estimates were averaged to reduce noise effect before performing equalization. 
\begin{figure}[!t]
	\centering
	
	\psfrag{-1.5}[c][c]{\small -$1.5$}
	\psfrag{-1}[c][c]{\small -$1$}
	\psfrag{-0.5}[c][c]{\small -$0.5$}
	\psfrag{0}[c][c]{\small $0$}
	\psfrag{0.5}[c][c]{\small $0.5$}
	\psfrag{1}[c][c]{\small $1$}
	\psfrag{1.5}[c][c]{\small $1.5$}
	
	\psfrag{I}{$I$}
	\psfrag{Q}{$Q$}
	
	\includegraphics[width=6cm]{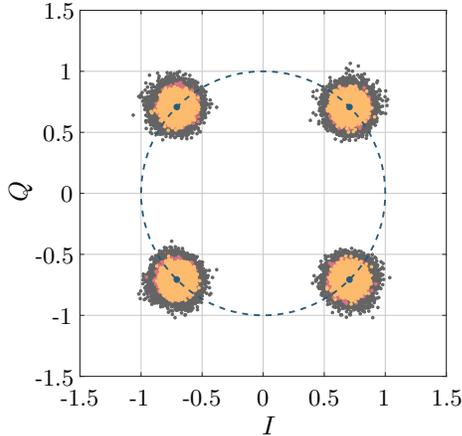}
	\caption{Normalized receive \ac{QPSK} constellations: sector-modulated \ac{OCDM}-based \ac{RadCom} system ({\color[rgb]{0.9882,0.7333,0.4275}$\CIRCLE$}), conventional \ac{OCDM}-based \ac{RadCom} system ({\color[rgb]{0.8471,0.4510,0.4980}$\CIRCLE$}), and \ac{OFDM}-based \ac{RadCom} system ({\color[rgb]{0.3922,0.3922,0.3922}$\CIRCLE$}). For reference, an unit-radius circle ({\color[rgb]{0.1490,0.3569,0.4824}\textbf{\textendash~\textendash}}) and a \ac{QPSK} constellation with unit symbol energy ({\color[rgb]{0.1490,0.3569,0.4824}$\CIRCLE$}) are shown.}\label{fig:RadCom_constellation}
	%\caption{Normalized receive \ac{QPSK} constellations: sector-modulated \ac{OCDM}-based \ac{RadCom} system ({\color[rgb]{0.9882,0.7333,0.4275}$\CIRCLE$}, estimated input \ac{SNR} of $\SI{29.65}{dB}$ and \ac{EVM} of $\SI{-26.27}{dB}$), conventional \ac{OCDM}-based \ac{RadCom} system ({\color[rgb]{0.8471,0.4510,0.4980}$\CIRCLE$}, estimated input \ac{SNR} of $\SI{29.74}{dB}$ and \ac{EVM} of $\SI{-26.20}{dB}$), and \ac{OFDM}-based \ac{RadCom} system ({\color[rgb]{0.3922,0.3922,0.3922}$\CIRCLE$}, estimated input \ac{SNR} of $\SI{29.65}{dB}$ and \ac{EVM} of $\SI{-27.64}{dB}$). For reference, an unit-radius circle ({\color[rgb]{0.1490,0.3569,0.4824}\textbf{\textendash~\textendash}}) and a \ac{QPSK} constellation with unit symbol energy ({\color[rgb]{0.1490,0.3569,0.4824}$\CIRCLE$}) are shown.}\label{fig:RadCom_constellation}
\end{figure}
The obtained normalized receive \ac{QPSK} constellations are shown in Fig.~\ref{fig:RadCom_constellation}. Based on the achieved results, both sector-modulated and conventional \ac{OCDM}-based \ac{RadCom} systems yield virtually the same communication performance. While the first has an estimate \ac{SNR} of $\SI{29.65}{dB}$ and an \ac{EVM} with mean value of $\SI{-26.27}{dB}$ and standard deviation of $\SI{5.59}{dB}$, an estimated input \ac{SNR} of $\SI{29.74}{dB}$ and \ac{EVM} with mean value of $\SI{-26.20}{dB}$ and standard deviation of $\SI{5.53}{dB}$ for the latter. As for the \ac{OFDM}-based \ac{RadCom} system, an estimated input \ac{SNR} of $\SI{29.65}{dB}$ and \ac{EVM} with mean value of $\SI{-27.64}{dB}$ and standard deviation of $\SI{6.30}{dB}$ were obtained. The higher mean and value and lower standard deviation  of the \ac{EVM} in the \ac{OCDM}-based systems can be explained by the fact that the degradation imposed by channel is spread over all subchirps. Conversely, while a lower \ac{EVM} mean value is experienced in the \ac{OFDM}-based \ac{RadCom} system, its higher standard deviation is due to different subcarriers experiencing different attenuation levels while propagating through the channel. More specifically, the subcarriers on the rightmost side of the spectrum experience around $\SI{10}{dB}$ more attenuation than the ones on the leftmost side (see Fig.~\ref{fig:H_CFR}), being therefore more severely degraded and having their receive \ac{QPSK} symbols shifted away from the reference constellation in Fig.~\ref{fig:RadCom_constellation}. This behavior agrees with the predicted higher robustness of the \ac{OCDM}-based \ac{RadCom} systems in frequency-selective channels, e.g., as reported in \cite{ouyang2016,moreira2021,omar2021,giroto2021_tmtt}, and leads to a better overall communication performance compared to their \ac{OFDM} counterpart.
%
%\corb{\cor{(GIVE DETAILS ON THE CONSIDERED CHANNEL)} the measured communication performance of the considered \ac{OCDM}-based \ac{RadCom} system is illustrated in Fig.~\ref{fig:RadCom_constellation}. In this figure, the estimated \ac{QPSK} constellation at the receiver side after equalization with the \acp{CIR} estimates contained in the $N_\mathrm{CP}$ subchirps of every \ac{OCDM} symbol is shown. For an \ac{SNR} level of \cor{$\SI{00}{dB}$} before communication processing, an error vector magnitude (EVM) of \cor{$\SI{00}{dB}$} is achieved.}

Finally, the simulated \acp{CCDF} of the baseband \ac{PAPR} for a single \ac{OCDM} symbol is shown for the investigated \ac{OCDM}-based radar system, its \ac{MU}/\ac{MIMO} variant, and the proposed sector-modulated \ac{OCDM}-based \ac{RadCom} system. For the sake of comparison, the \acp{PAPR} of considered \ac{OFDM}- and conventional \ac{OCDM}-based \ac{RadCom} systems in the previous radar and communication analysis are also shown as a baseline. As in \cite{giroto2021_tmtt}, the \ac{PAPR} was calculated assuming an oversampling factor of $20$ to capture fast variations of time-domain signals. The particularly low baseband \ac{PAPR} for the considered \ac{OCDM}-based radar system and its \ac{MU}/\ac{MIMO} variant is explained by the single active subchirp in \eqref{eq:FresnelPilot} and \eqref{eq:FresnelPilotMIMO}, which yields a complex exponential signal with a virtually constant envelop in the discrete-time domain after the \ac{IDFnT} in \eqref{eq:IDFnT}. As for the sector-modulated \ac{OCDM}-based \ac{RadCom} system, the presence of active, modulated subchirps in \eqref{eq:X_frame_RadCom} yields nearly $\SI{6}{dB}$ higher average \ac{PAPR}. Compared to the \ac{OFDM}- and conventional \ac{OCDM}-based \ac{RadCom} systems, which perform equally, the sector-modulated \ac{OCDM}-based \ac{RadCom} system presents reduced \ac{PAPR} by around $\SI{1.1}{dB}$. If lower communication data rates are aimed, this difference can be increased by deactivating some of the \ac{QPSK}-modulated subchirps, which yields a similar effect to the \ac{PAPR} reduction by tone reservation in \ac{OFDM}-based systems.%result in lower \ac{PA} amplifier requirements in a practical system design.}

\begin{figure}[!t]
	\centering
	
	\psfrag{AAAA}[c][c]{\small $0$}
	\psfrag{BBBB}[c][c]{\small $0.2$}
	\psfrag{CCCC}[c][c]{\small $0.4$}
	\psfrag{DDDD}[c][c]{\small $0.6$}
	\psfrag{EEEE}[c][c]{\small $0.8$}
	\psfrag{FFFF}[c][c]{\small $1$}
	
	\psfrag{AAA}[c][c]{\small $0$}
	\psfrag{BBB}[c][c]{\small $4.5$}
	\psfrag{CCC}[c][c]{\small $9$}
	\psfrag{DDD}[c][c]{\small $13.5$}
	\psfrag{EEE}[c][c]{\small $18$}
	
	\psfrag{X}{$x$}
	\psfrag{YYYYYYYYYYYYY}{$P(\mathit{PAPR}_\mathrm{dB}>x)$}
	
	\includegraphics[width=6cm]{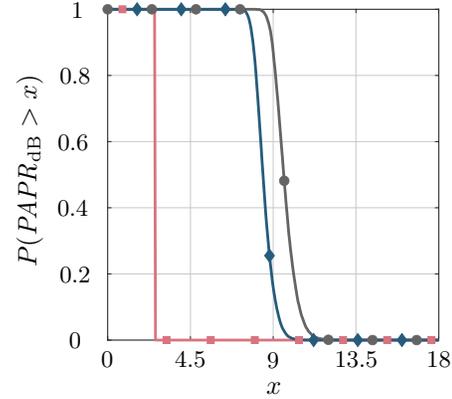}
	\caption{Simulated baseband PAPR for the considered OCDM-based radar system and its MU/MIMO variant per transmit channel ({\color[rgb]{0.8471,0.4510,0.4980}$\blacksquare$}), sector-modulated OCDM-based RadCom system ({\color[rgb]{0.1490,0.3569,0.4824}$\blacklozenge$}) and an OFDM- and conventional OCDM-based RadCom systems ({\color[rgb]{0.3922,0.3922,0.3922}$\CIRCLE$}), all with symbol length of $N=2048$.}\label{fig:PAPR}
\end{figure}

%***********************************************************%
%-----------------------------------------------------------%
%***********************************************************%

\section{Conclusion}\label{sec:conclusion}

This article has investigated the use of discrete-Fresnel domain channel estimation for \ac{OCDM}-based radar systems. The proposed processing strategy relies on a strategic subchirp allocation for \ac{OCDM} symbols in the discrete-Fresnel domain and exploits the pulse-compression-like effect of the \ac{DFnT}. After a thorough mathematical formulation of the individual and joint effects of time and frequency shifts on \ac{OCDM} signals in the discrete-Fresnel domain, discrete-Fresnel domain radar channel estimation strategies have been proposed for both \ac{SISO} and \ac{MIMO} radar applications, as well as for \ac{RadCom} operation. Finally, a detailed performance analysis supported by simulation and measurements with a \ac{RTS} has been carried out assuming a mid-range \ac{HAD} scenario to validate the contributions of this article and compare the achieved performances with the ones of \ac{OFDM}- and conventional \ac{OCDM}-based radar/\ac{RadCom} systems.

The achieved results have shown that the investigated discrete-Fresnel domain channel estimation for \ac{OCDM}-based radar and \ac{RadCom} systems yields comparable performance to \ac{OFDM}-based radar/\ac{RadCom} systems in all its forms, i.e., \ac{SISO} and \ac{MIMO} operations or sector-modulated \ac{RadCom} operation. Significant differences, namely peak splitting and range migration, are only observed for Doppler shifts associated with relative radial velocities that are not to be expected in practice. Compared to the conventional \ac{OCDM}-based \ac{RadCom} system from a previous work, all of the proposed strategies yield improved radar sensing performance as their range sidelobes are not influenced by modulated data symbols onto subchirps. Additionally, the proposed strategies for \ac{SISO} and \ac{MIMO} \ac{OCDM}-based radar sensing yield significantly lower \ac{PAPR} than their \ac{OFDM} and conventional \ac{OCDM} counterpart. As for the sector-modulated \ac{OCDM}-based \ac{RadCom} system, the experienced \ac{PAPR} reduction is lower, but can be improved in exchange for data rate reduction. For radar-centric systems, where very high data rates are not necessarily a requirement, the proposed sector-modulated \ac{OCDM}-based \ac{RadCom} system also appears as an attractive alternative to its \ac{OFDM} and conventional \ac{OCDM} counterparts, since it shows robustness for both radar sensing and communication.

\appendix
\renewcommand{\theequation}{A.\arabic{equation}}
\setcounter{equation}{0}
%
%\textbf{IDFnT}
%
%\setcounter{equation}{0}
%\begin{equation}\label{eq:IDFnT}
%x_n = \frac{1}{N}\e^{\im\frac{\pi}{4}}\sum\limits_{k=0}^{N-1} \dot{X}_k~\e^{-\im\frac{\pi}{N}(n-k)^2}
%\end{equation}
%
%\textbf{DFnT}
%
%\begin{equation}\label{eq:DFnT}
%\dot{X}_k = \e^{-\im\frac{\pi}{4}}\sum\limits_{n=0}^{N-1} x_n~\e^{\im\frac{\pi}{N}(n-k)^2}
%\end{equation}
%
%\textbf{Finite geometric series}
%
The finite geometric series and its solution used for the derivation of equations in Sections~\ref{sec:time-freqConv} and \ref{sec:channel_est} are expressed as
\begin{eqnarray}\label{eq:geoSeries}
	\sum\limits_{\eta=0}^{\gamma}\alpha^\eta &=& 1 + \alpha + \alpha^2 + \cdots + \alpha^\gamma\nonumber\\
	&=& \frac{\alpha^{\gamma+1}-1}{\alpha-1}\quad\text{for}\quad \alpha\neq1
\end{eqnarray}
for $\eta\in\mathbb{N}$, $\gamma\in\mathbb{N}$, $\alpha\in\mathbb{C}$.

\bibliographystyle{IEEEtran}
\bibliography{OCDMchannelEstimation}

%\newpage
%.
%\newpage
%
\begin{IEEEbiography}[{\includegraphics[width=1in,height=1.25in,clip,keepaspectratio]{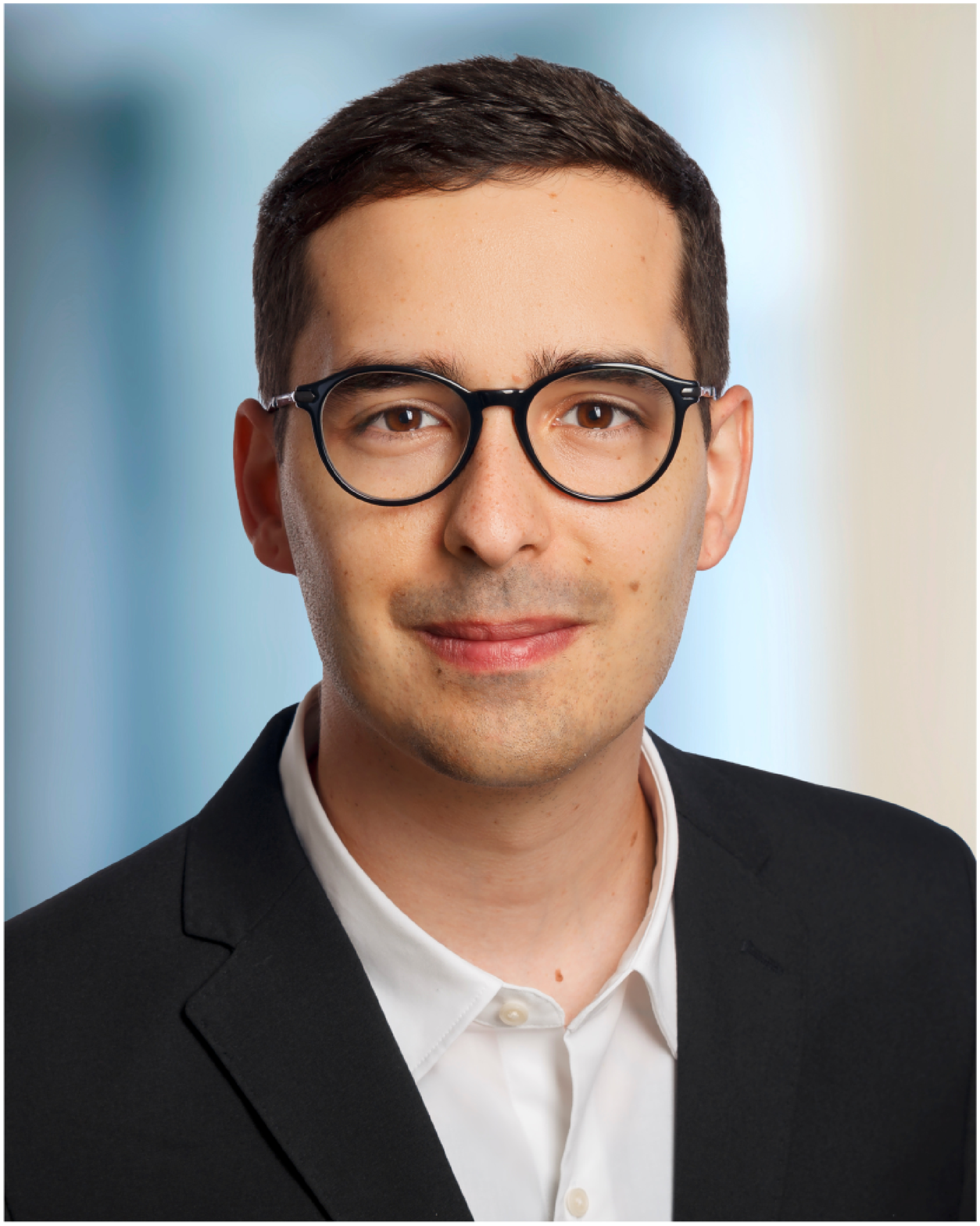}}]{Lucas Giroto de Oliveira}
	(Graduate Student Member, IEEE) received the B.Sc. and M.Sc. degrees in electrical engineering with a major in electronic systems from the Federal University of Juiz de Fora (UFJF), Juiz de Fora, Brazil, in 2017 and 2019, respectively. He is currently pursuing the Dr.-Ing. (Ph.D.E.E.) degree at the Karlsruhe Institute of Technology (KIT), Karlsruhe, Germany.
	
	He was with the Laboratório de Comunicações (LCom), UFJF, from June 2014 to March 2019. From April 2015 to March 2016, he was also with the Digital Communications Group, University of Duisburg-Essen, Duisburg, Germany. Since April 2019, he has been with the Institute of Radio Frequency Engineering and Electronics (IHE), KIT, where he is currently a Research Associate. His research interests are in the areas of signal processing, digital communication, and their applications to integrated radar sensing and communication systems and networks.
	
	Mr. Giroto de Oliveira was a recipient of the Science without Borders (CsF) scholarship funded by Coordenação de Aperfeiçoamento de Pessoal de Nível Superior (CAPES), Brazil, from September 2014 to March 2016, and of the Research Grant for Doctoral Programmes in Germany from the German Academic Exchange Service (DAAD), Germany, from April 2019 to March 2021. He was also a co-author of the Best Conference Paper award winning paper at the 2022 International Workshop on Antenna Technology (iWAT).
	%Mr. Giroto de Oliveira was a recipient of the Science without Borders (CsF) scholarship funded by Coordenação de Aperfeiçoamento de Pessoal de Nível Superior (CAPES), Brazil, from September 2014 to March 2016. He also received the Research Grant for Doctoral Programmes in Germany from the German Academic Exchange Service (DAAD), Germany, from April 2019 to March 2021. 
\end{IEEEbiography}

\begin{IEEEbiography}[{\includegraphics[width=1in,height=1.25in,clip,keepaspectratio]{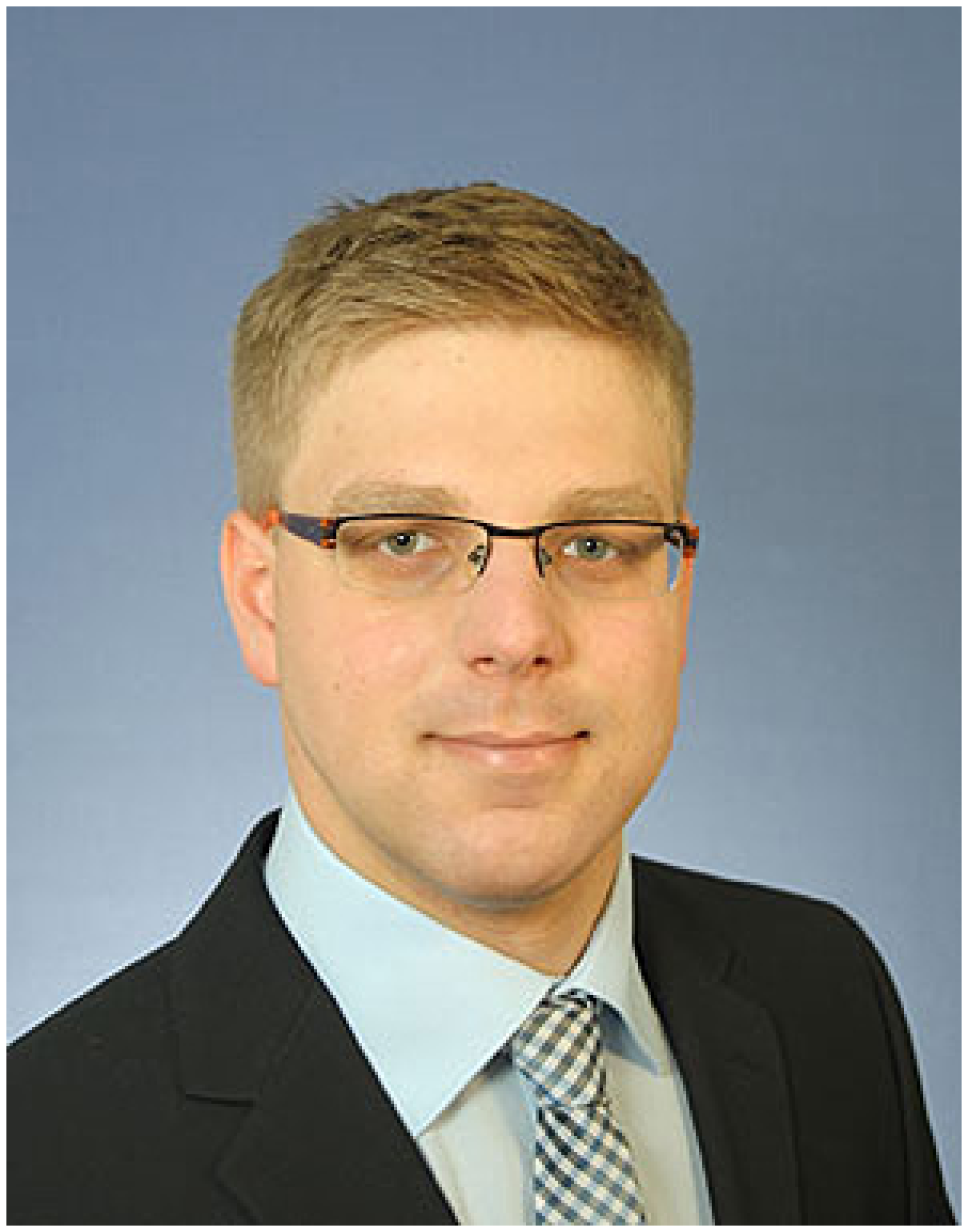}}]{Benjamin Nuss}
	(Graduate Student Member, IEEE) received the B.Sc. and M.Sc. degrees in electrical engineering and information technology from the Karlsruhe Institute of Technology (KIT), Karlsruhe,	Germany, in 2012 and 2015, respectively, and the Dr.-Ing. (Ph.D.E.E) degree from KIT, in 2021.
	
	He is currently working as a Group Leader for radar systems at the Institute of Radio Frequency Engineering and Electronics (IHE). The focus of his work is on the development of efficient future radar waveforms and interference mitigation techniques for multi-user scenarios. His current research interests include orthogonal
	frequency-division multiplexing-based multiple-input multiple-output radar systems for future automotive applications and drone detection. He was also the author and co-author of the best publication of 2021 and 2022 within the ITG Society, which was awarded the VDE ITG Prize.
\end{IEEEbiography}

\begin{IEEEbiography}[{\includegraphics[width=1in,height=1.25in,clip,keepaspectratio]{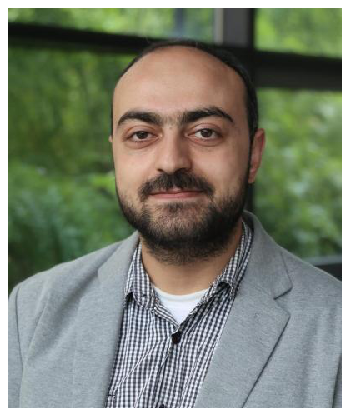}}]{Mohamad Basim Alabd}
	(Graduate Student Member, IEEE) received the B.Sc. and M.Sc. degrees in electronics and telecommunications from Al-Baath University, Homs, Syria, in 2010 and 2014, respectively.
	
	He was a Research Associate with the Chair of Information Systems, Innovation \& Value Creation	(WI1), Friedrich-Alexander-Universität Erlangen-N\"urnberg, Nuremberg, Germany. From 2010 to 2014, he was with Syrian Telecommunication Company, Homs, and worked there in different positions. He has been a Research Associate with the Institute of Radio Frequency Engineering and Electronics (IHE), Karlsruhe Institute of Technology, Karlsruhe, Germany, since 2017. His current research interests include chirp sequence (CS) joint radar-communication and orthogonal frequency-division multiplexing-based multiple-input multiple-output radar systems for future automotive applications.
\end{IEEEbiography}

\begin{IEEEbiography}[{\includegraphics[width=1in,height=1.25in,clip,keepaspectratio]{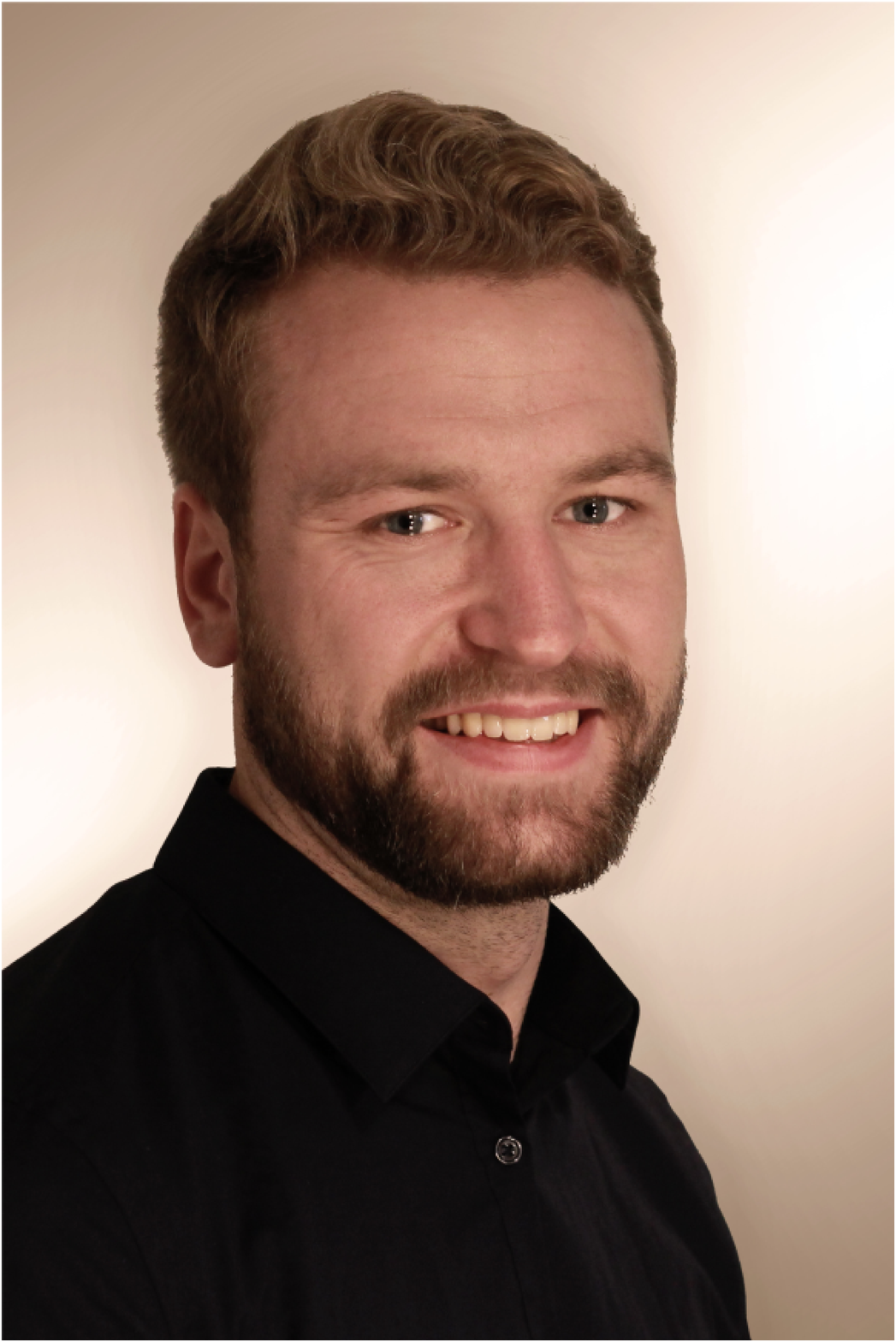}}]{Axel Diewald}
	(Graduate Student Member, IEEE) received the B.Sc. and M.Sc. degrees in electrical engineering and information technology from the Karlsruhe Institute of Technology (KIT), Karlsruhe, Germany, in 2015 and 2017, respectively, where he is currently pursuing the Ph.D. degree in electrical engineering (EE) at the Institute of Radio Frequency Engineering and Electronics (IHE).
	
	His main research interests include digital radar target simulation for the purpose of automotive radar sensor validation and realistic target modeling.
\end{IEEEbiography}

\begin{IEEEbiography}[{\includegraphics[width=1in,height=1.25in,clip,keepaspectratio]{./Photos/YuehengLi}}]{Yueheng Li}
	(Graduate Student Member, IEEE) received his B.Sc. in telecommunication science and technology from Shandong University, Jinan, China in 2014. In 2018, he further received his M.Sc. in communication and information technology from University of Bremen, Bremen, Germany. Since 2018, he has been a scientific employee and working toward the Dr.-Ing. (Ph.D.E.E) degree at the Institute of Radio Frequency Engineering and Electronics (IHE), Karlsruhe, Germany. His research focuses on mobile wireless communication systems using programmable metasurface.
\end{IEEEbiography}

\begin{IEEEbiography}[{\includegraphics[width=1in,height=1.25in,clip,keepaspectratio]{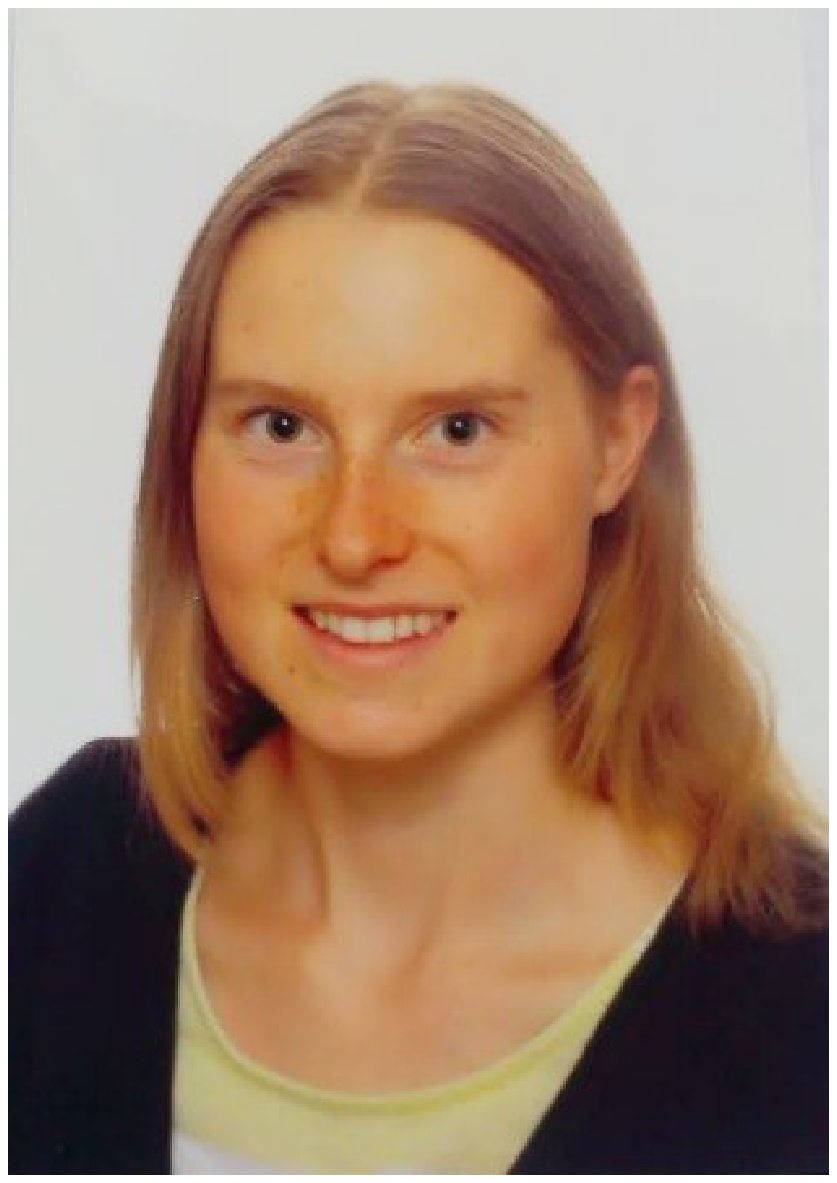}}]{Linda Gehre}
	%received the B.Sc. degree in electrical engineering and information technology from the Karlsruhe Institute of Technology (KIT), Karlsruhe, Germany, in 2021, where she is currently working on her M.Sc. degree. At the same time, she is working as a student assistant at the Institute of Radio Frequency Engineering and Electronics (IHE) in the field of radar sensing and communication systems.
	received the B.Sc. degree in electrical engineering and information technology from the Karlsruhe Institute of Technology (KIT), Karlsruhe, Germany, in 2021, where she is pursuing the M.Sc. degree and working as a student assistant at the Institute of Radio Frequency Engineering and Electronics (IHE). Her research interests are in the field of radar sensing and communication systems.
\end{IEEEbiography}

\begin{IEEEbiography}[{\includegraphics[width=1in,height=1.25in,clip,keepaspectratio]{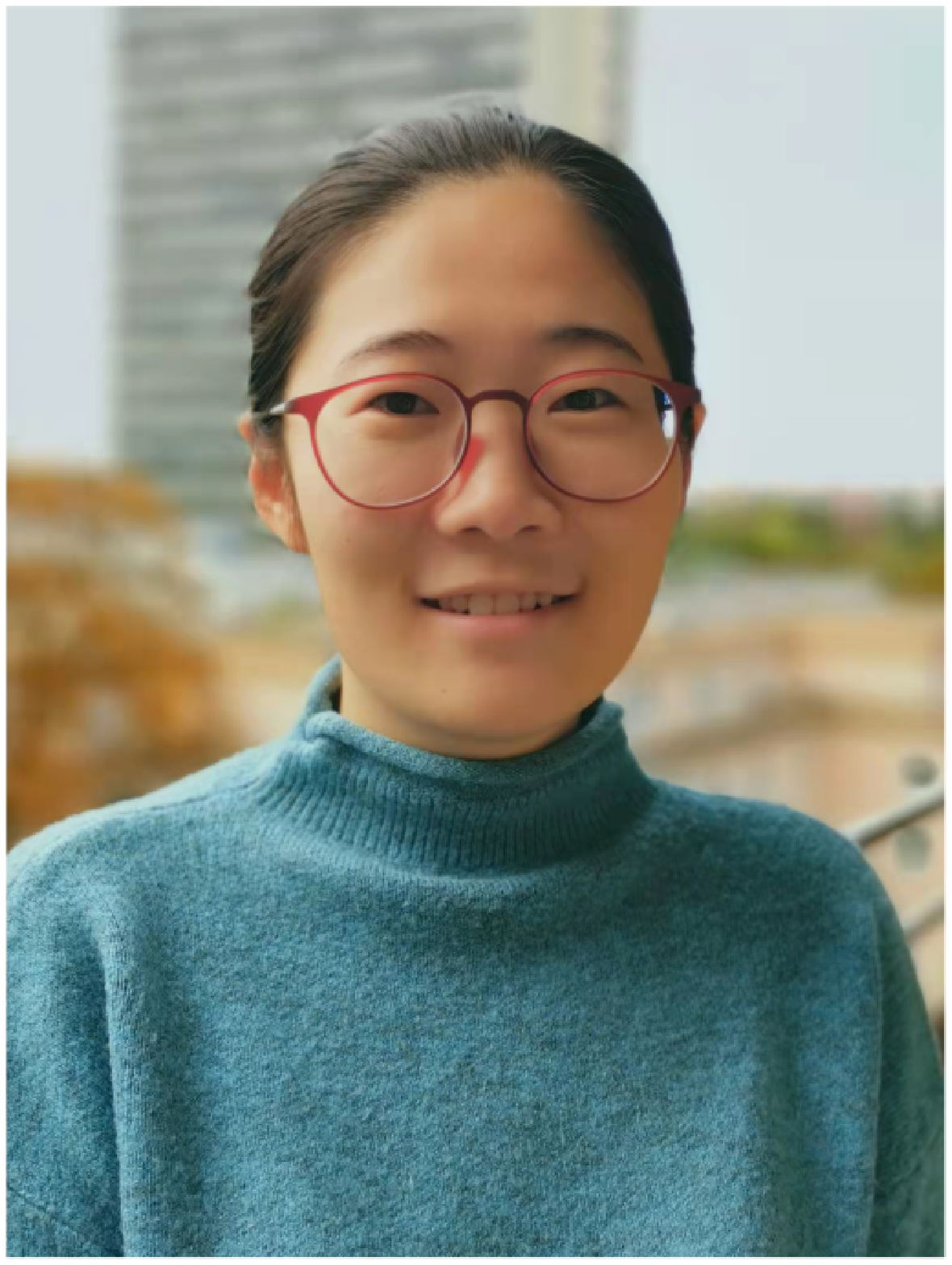}}]{Xueyun Long}
	received the B.Sc. degree in electronic information engineering from the Beijing Institute of Technology, Beijing, China, in 2018, the M.Sc. degree in electronic and information technology, in 2021, from the Karlsruhe Institut of Technology, Karlsruhe, Germany, where she is currently working toward the Dr.-Ing. (Ph.D.E.E) degree in electronical engineering with the Institute of Radio Frequency Engineering and Electronics, Karlsruhe, Germany. Her research interests include exposure and programmable metasurface.
\end{IEEEbiography}

\begin{IEEEbiography}[{\includegraphics[width=1in,height=1.25in,clip,keepaspectratio]{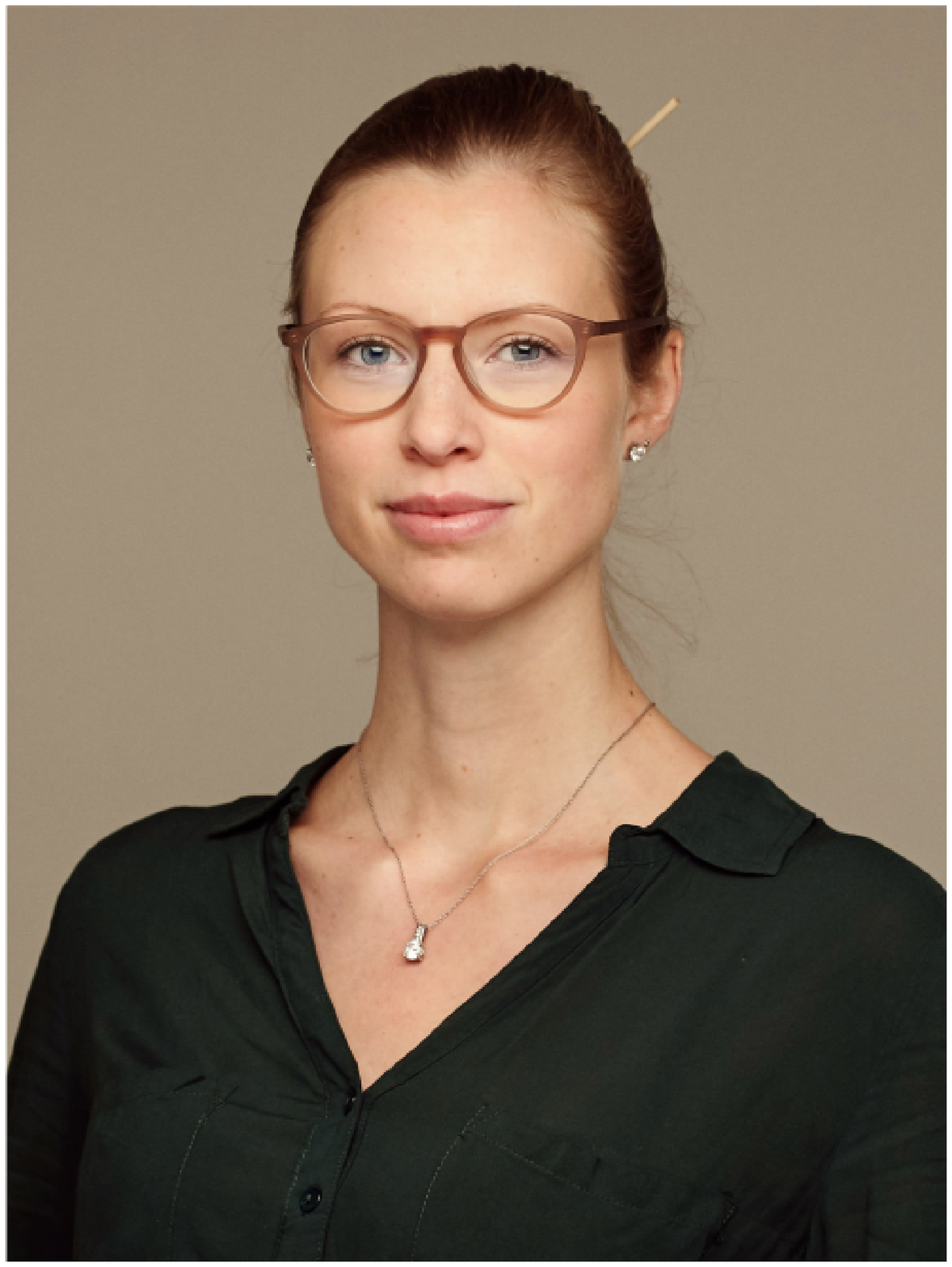}}]{Theresa Antes}
	received the B.Sc. and M.Sc. degrees in electrical engineering and information technology from the Karlsruhe Institute of Technology (KIT), Karlsruhe, Germany, in 2018 and 2020, respectively, where she is currently pursuing the Ph.D. degree in electrical engineering (EE) at the Institute of Radio Frequency Engineering and Electronics (IHE). Her main research interests include signal processing, high-accuracy radar systems, and radar-based gesture recognition.
\end{IEEEbiography}

\begin{IEEEbiography}[{\includegraphics[width=1in,height=1.25in,clip,keepaspectratio]{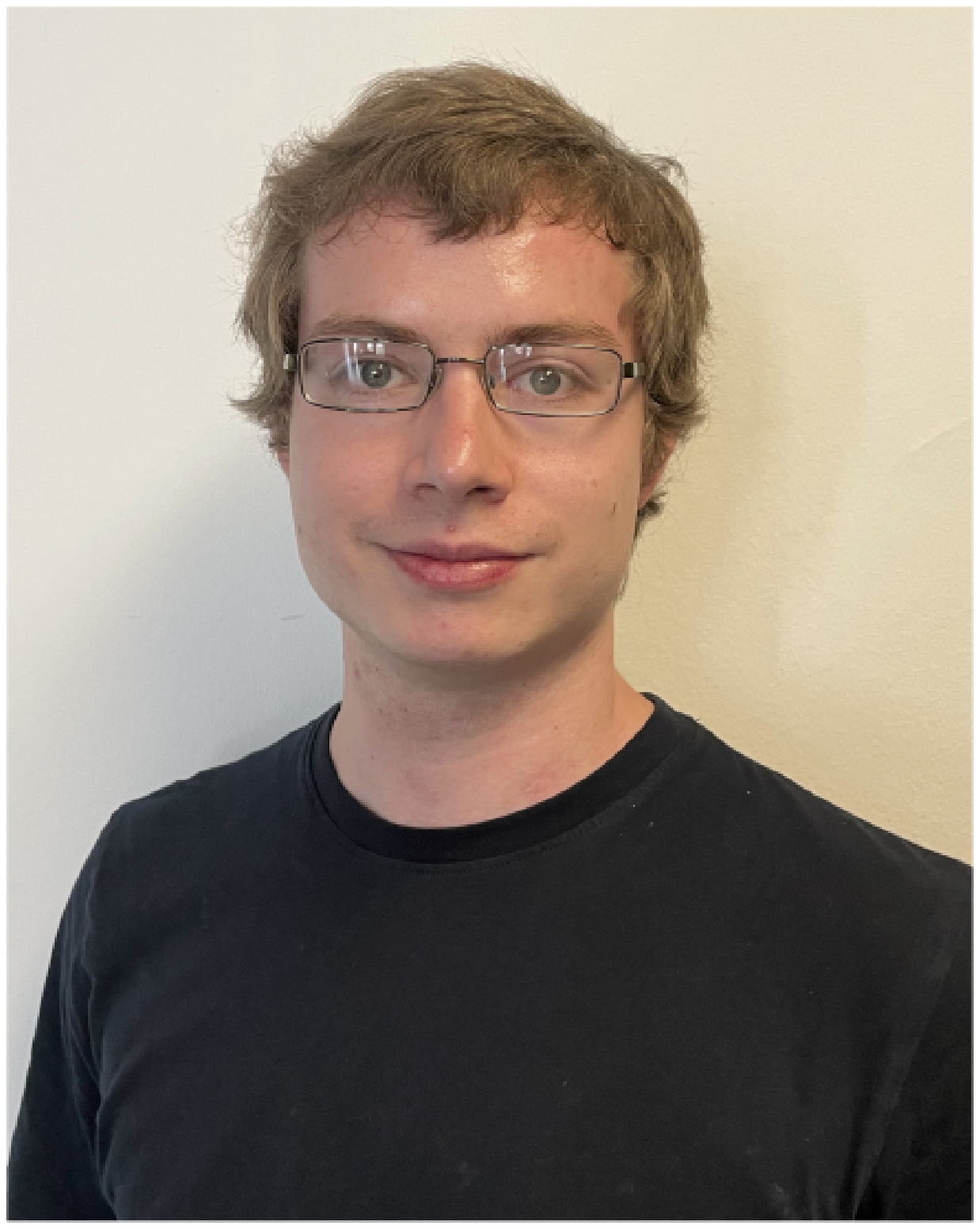}}]{Johannes Galinsky}
	received the B.Sc. and M.Sc. degrees in electrical engineering and information technology from the	Karlsruhe Institute of Technology (KIT), Karlsruhe, Germany, in 2019 and 2022, respectively, where he
	is currently pursuing the Ph.D. degree in electrical engineering (EE) at the Institute of Radio Frequency Engineering and Electronics (IHE). His main research interests include signal processing, digital radar and joint communication \& sensing.
\end{IEEEbiography}

\vfill

\begin{IEEEbiography}[{\includegraphics[width=1in,height=1.25in,clip,keepaspectratio]{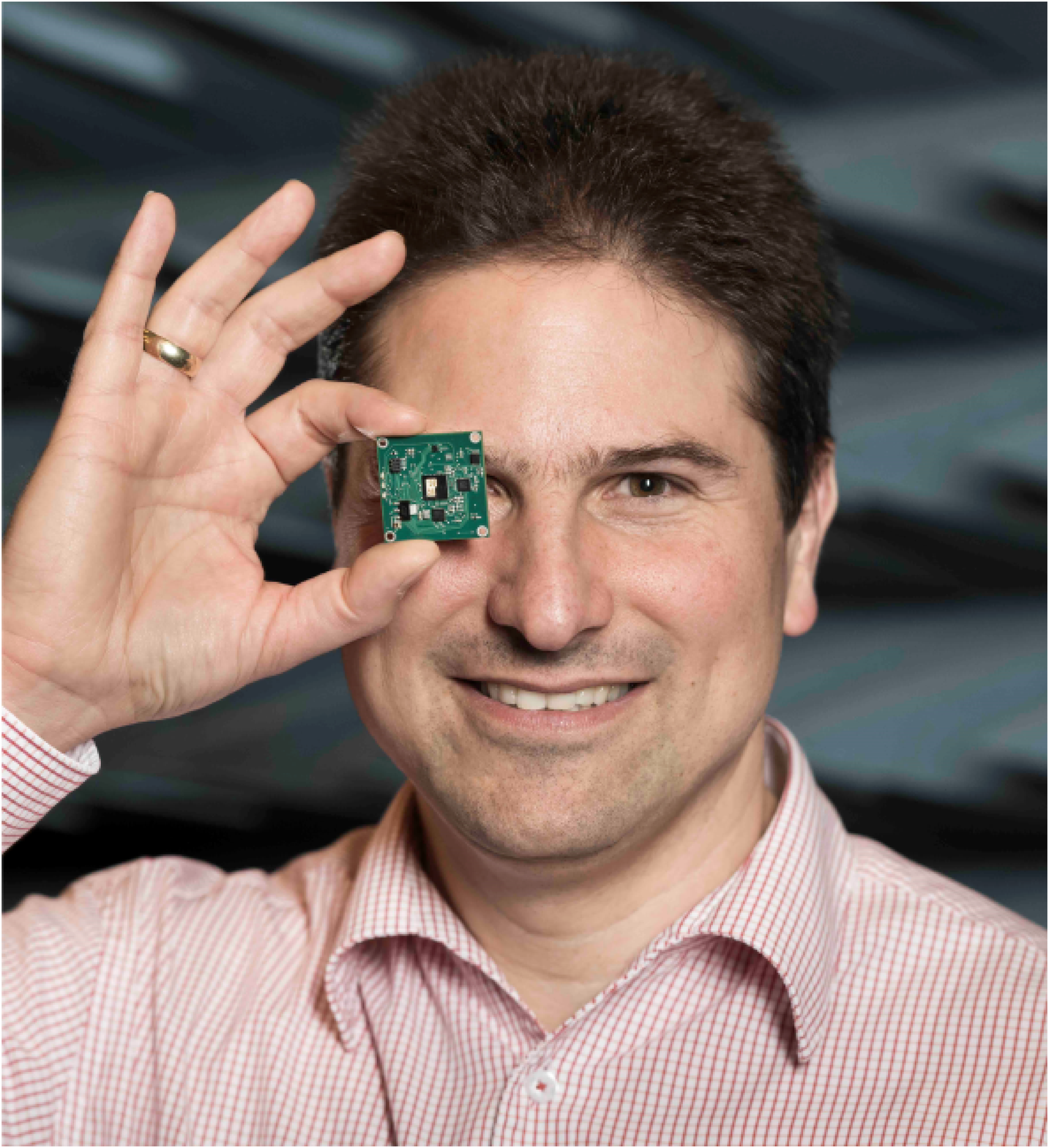}}]{Thomas Zwick}
	(Fellow, IEEE) received the	Dipl.-Ing. (M.S.E.E.) and Dr.-Ing. (Ph.D.E.E.) degrees from the Universität Karlsruhe (TH), Karlsruhe, Germany, in 1994 and 1999, respectively.
	
	From 1994 to 2001, he was a Research Assistant with the Institut f\"ur H\"ochstfrequenztechnik und Elektronik (IHE), TH. In February 2001, he joined IBM as Research Staff Member at the IBM Thomas J. Watson Research Center, Yorktown Heights, NY, USA. From October 2004 to September 2007, he was with Siemens AG, Lindau, Germany. During that period, he managed the RF development team for automotive radars. In October 2007, he became a Full Professor with the Karlsruhe Institute of Technology (KIT), Karlsruhe. He is currently the Director of the Institute of Radio Frequency Engineering and Electronics (IHE), KIT. He is a co-editor
	of three books and the author or a co-author of 120 journal articles, over 400 contributions at international conferences, and 15 granted patents. His research topics include wave propagation, stochastic channel modeling, channel measurement techniques, material measurements, microwave techniques, millimeter-wave antenna design, wireless communication, and radar system design.
	
	Dr. Zwick has been a member of the Heidelberg Academy of Sciences and Humanities since 2017. Since 2019, he has been a member of acatech (German National Academy of Science and Engineering). His research team received over ten best paper awards at international conferences. In 2013, he was the General Chair of the International Workshop on Antenna Technology (iWAT 2013) in Karlsruhe and the IEEE MTT-S International Conference on Microwaves for Intelligent Mobility (ICMIM) in Heidelberg in 2015. He was	the TPC Chair of the European Microwave Conference (EuMC) 2013 and the General TPC Chair of the European Microwave Week (EuMW) 2017.
	In 2023, he will be the General Chair of EuMW in Berlin. He has served on the technical program committees (TPC) of several scientific conferences. From 2008 to 2015, he was the president of the Institute for Microwaves and Antennas (IMA). He was selected as a Distinguished IEEE Microwave Lecturer for the 2013-2015 period with his lecture on ``QFN Based Packaging	Concepts for Millimeter-Wave Transceivers.'' In 2019, he became the Editor-in-Chief of the \textsc{IEEE Microwave and Wireless Components Letters}. In 2022 he was awarded an honorary doctorate of the Faculty of Electrical Engineering and Informatics at the Budapest University of Technology and Economics in Hungary.
\end{IEEEbiography}

\end{document}